\documentclass[conference]{IEEEtran}
\IEEEoverridecommandlockouts
\usepackage{hyperref}
\usepackage{amsmath}
\usepackage{amssymb}
\usepackage{graphicx}
\usepackage{bm}
\usepackage{paralist}
\usepackage{subfigure}
\usepackage{color}
\usepackage{array}
\usepackage{cite}
\usepackage{times}
\usepackage{setspace}
\usepackage{rotating}
\usepackage{pdflscape}
\usepackage{url}
\usepackage{balance}

\allowdisplaybreaks

\newcolumntype{C}[1]{>{\centering\let\newline\\\arraybackslash\hspace{0pt}}m{#1}}

\newcommand{\nC}{|{\cal C}|}

\newboolean{onecol}
\setboolean{onecol}{false}

\newcommand{\specialcell}[2][c]{%
  \begin{tabular}[#1]{@{}c@{}}#2\end{tabular}}

\newcommand{\sbt}{\,\begin{picture}(-1,1)(-1,-3)\circle*{3}\end{picture}\ }

\makeatletter
\def\ps@headings{%
\def\@oddhead{\mbox{}\scriptsize\rightmark \hfil \thepage}%
\def\@evenhead{\scriptsize\thepage \hfil \leftmark\mbox{}}%
\def\@oddfoot{}%
\def\@evenfoot{}}
\makeatother
\newcommand{\set}[1]{\ensuremath{\mathcal #1}}

\newcommand{\separator}{
  \begin{center}
    \rule{\columnwidth}{0.3mm}
  \end{center}
}

\def\lf{\left}
\def\ri{\right}

\newtheorem{theorem}{Theorem}[section]

\newcommand{\expect}[1]{\mathbb{E}[ #1 ]}

\newcommand{\bbexpect}[1]{\mathbb{E}\lf[ #1 \ri]}

\newcommand{\beq}{\begin{eqnarray*}}
\newcommand{\eeq}{\end{eqnarray*}}
\newcommand{\beqn}{\begin{eqnarray}}
\newcommand{\eeqn}{\end{eqnarray}}
\newcommand{\bemn}{\begin{multiline}}
\newcommand{\eemn}{\end{multiline}}

\newcommand{\sqeq}{\addtolength{\thinmuskip}{-4mu}
\addtolength{\medmuskip}{-4mu}\addtolength{\thickmuskip}{-4mu}}

\newcommand{\unsqeq}{\addtolength{\thinmuskip}{+4mu}
\addtolength{\medmuskip}{+4mu}\addtolength{\thickmuskip}{+4mu}}

\renewcommand{\baselinestretch}{0.98}

\IEEEoverridecommandlockouts

\begin{document}
%
\title{On the Delay Scaling Laws of Cache Networks}

\author
{
  \IEEEauthorblockN{Boram Jin, Deawoo Kim, \\ Jinwoo Shin, and Yung Yi}
  \IEEEauthorblockA{EE Dept., KAIST, Korea \\
  }
  \and
  \IEEEauthorblockN{Se-Young Yun}
  \IEEEauthorblockA{School of EE, KTH, \\ Stockholm, Sweden
  }
  \and
  \IEEEauthorblockN{Seongik Hong and Byoung-Joon (BJ) Lee}
  \IEEEauthorblockA{Samsung Advanced Institute of Tech., \\ Suwon, Korea
  }
}

\maketitle

\begin{abstract}
  The Internet is becoming more and more content-oriented. CDN (Content
  Distribution Networks) has been a popular architecture compatible with
  the current Internet, and a new revolutionary paradigm such as ICN (Information
  Centric Networking), has also been extensively studied as a concept
  for several next-generation Internet  architectures. One of main components
  in both CDN and ICN is network caching. 
  Despite a surge of extensive use of network cashing in the current and
  future Internet architectures, analysis on the performance of cache
  networks in terms of their network topologies and content popularities
  are still quite limited due to complex inter-plays among various
  components and thus analytical intractability.

  In this paper, we study asymptotic delay performance of cache
  networks, in particular, focusing on the impact of heterogeneous
  content popularities and nodes' geometric `importances' in caching
  policy.  First, we consider the case when per-node cache size is
  homogenous (i.e., uniform), and provide precise delay scaling laws in
  terms of several factors including content popularities, content
  placing policies, cache budgets, average routing distances and numbers
  of contents. The important characteristic of our scaling laws is {\em
    topology-independence}, i.e., the asymptotic delay performance
  depends just on the distribution or the average of request routing
  distance, independent from other network topological
  properties. Second, we study the joint impact of
  popularity-based content placements and heterogeneous (i.e.,
  non-uniform) cache sizes which depend on the geometric importances of
  nodes. We show that the asymptotic delay performance of cache networks
  can be significantly reduced by heterogeneous cache sizing if the
  network has a (almost) regular spanning tree. 
We draw various engineering implications from our
  analysis and perform extensive simulations on the real Internet topology.
\end{abstract}


\section{Introduction}

\subsection{Motivation}
Due to a recent shift that the Internet has increasingly become
content-delivery oriented, Internet researchers constantly seek for ways
of adapting the Internet to such a shift, e.g., advancing CDN (Content
Distribution Networks) technologies as an evolutionary approach, or 
proposing revolutionary architectures such as ICN (Information Centric
Networking) and CCN (Content Centric Networking)
\cite{DONA,NDN,CCN,4WARD,PSIRP/PURSUIT,COMET}. IP (Internet Protocol) was designed simply for
host-to-host conversation, giving rise to a mismatch with content-based
delivery, regarded as the root cause of several fundamental
problems of the current Internet, e.g., security and mobility.  ICN/CCN
proposes to change the basic abstraction of the Internet from
“end-to-end delivery of packets to named hosts” to “secure
dissemination of named contents.”



In a content-oriented architecture (whether it is evolutionary or
revolutionary), content caching seems to be a crucial component to
reduce delay of fetching contents and/or overall traffic transport
cost, often forming a group of large-scale caches, namely a cache
network. While networked caches have already appeared in the past, e.g.,
web caches \cite{che2001analysis, che2002hierarchical}, they were mainly
small-scale ones based on simple (e.g., hierarchical) topological
structures. Despite an array of recent research interests in content
search, scalable architecture, and performance analysis in cache
networks (see Section~\ref{sec:related} for details), analytically
understanding networked caches is known to be a daunting task in general, which leaves much to be researched in the upcoming
years. The main challenge 
comes from complex inter-plays among several components such as content
request routing, network topology, and heterogeneous per-cache
budget, and dynamic cache replacement policies such as LFU (Least-Frequently-Used)
and LRU (Least-Recently-Used).
For example, understanding cache replacement policies even just for a
{\em single} cache is known to be challenging
\cite{dan1990approximate,jelenkovic1999asymptotic}.




\subsection{Our Contribution}

In this paper, we perform asymptotic delay analysis of large-scale cache networks. Our focus lies in quantitively
understanding the relation between content popularity and delay
performance as well as the impact of heterogeneity in terms of ``nodes geometric
importance'' (i.e., more caching at caches with larger accessibility).

For mathematical tractability, we consider {\em static cache policies},
where contents are placed in caches in the networks apriori and they
are not replaced over time.  This removes the need of considering
dynamic cache replacements which is regarded as one of the most complex
parts in cache network analysis. However, it does not incur too much
loss of generality since static cache policies can be regarded as
approximation schemes or steady-state regimes of dynamic, general ones
(see Section \ref{sec:simulation} for simulation based validations). For
example, a random, dynamic replacement policy can be approximated by a
static, random content placement on caches. A better replacement policy
would exploit temporal locality whose examples include LFU
and LRU. Those policies
can be approximately captured by static content placements which
consider popularities in content requests, e.g., placing more popular
contents in a cache with higher priority. This static policy has also
been adopted in other cache network analysis research
\cite{psaras2011modelling,gitzenis2012asymptotic}. 
In what follows, we
summarize our key contributions.

\smallskip
\begin{compactenum}[\bf \em C1]
\item {\bf \em Delay scaling laws: Homogenous per-node cache size.}	
  We first study the asymptotic delay of cache networks
  under homogenous per-node cache sizes, i.e., cache sizes are uniform
  among nodes.  To this end, we develop an analysis module that provides
  the expected delay for a given routing distance between a content requester
  and the original content server, where the expectation is taken over
  (potentially) random server locations and random content placement policies.
  We asymptotically study the module for representative content
  placement policies, ranging from random to
  popularity-driven. This module is a highly versatile `black-box'
  tool, enabling us to study the asymptotic delay performance
  independent from the details of cache network topology and request routing policy.
  Our results reveal precise asymptotic performances of cache networks
  in terms of content popularities, content
  placement policies, cache sizes and number of contents, which guides
  to design an efficient cache network in real-world scenarios. 

\smallskip
\item {\bf \em Delay scaling laws: Heterogeneous per-node cache size.}
  Second, we study the asymptotic delay under heterogeneous per-node cache sizes, i.e., cache sizes are
  possibly non-uniform among nodes.  The heterogeneity makes the
  analysis much challenging since the `topology-independent' module we
  developed in the homogenous case {\bf \em C1} is no longer applicable, i.e., one
  has to resolve a non-trivial geometric coupling between heterogeneous
  per-node cache sizes and content popularities. Due to
  such technical hardness, we focus on special network topologies where
  request routing policies consist of shortest-paths on a regular
  spanning tree, and obtain corresponding delay scaling laws under a simple sizing policy that has more cache sizes at geometrically important nodes.
  Our scaling laws imply that the caching gain incurred by
  heterogeneous cache sizing using nodes' geometric importances
  increases; $\log$-order delay reduction over homogeneous caching
  sizes.  
\end{compactenum}

We also provide simulation results to validate our theoretical results in {\bf \em C1}
and {\bf \em C2}. In our results, we consider dynamic content request
arrivals and run various, dynamic cache replacement policies: random,
LFU, and LRU, and observe that our theoretical analysis have good matches
with the results from simulations. 


\section{Related Work}
\label{sec:related}

Analyzing cache performance started from a single-scale case
\cite{jelenkovic1999asymptotic, jelenkovic2005near,dan1990approximate},
where the main focus was on deriving asymptotic or approximate
closed-form of cache hit probabilities for well-known cache policies such
as LRU, LFU, and FIFO, often on the assumption of IRM (Independence
Reference Model) (no temporal correlations in the stream of requests) for tractability. A network of cache, in spite of only for limited topology, has been
studied for web caches.  The work \cite{che2001analysis,che2002hierarchical} adopted a fluid model for analyzing a hierarchical caching (i.e., caching at clients,
institutions, regions, etc.), and proposed a new, analysis-inspired
caching scheme. The authors in \cite{rodriguez2001analysis} studied
tradeoffs between hierarchical and distributed caching (i.e., caching at
only institutional edges), and proposed a hybrid scheme that has lower
latency. Recently, the work
\cite{fricker2012versatile} mathematically explained the intuitions in
\cite{che2001analysis}.

Over the recent three years, mainly due to emergence of
information-centric networking, extensive attention has been made to
analysis of general cache networks. The authors in
\cite{borst2010distributed} developed cooperative cache management schemes inspired by a
solution of a linear programming (with objective of maximizing the
traffic volume served by caches) over a tree-based cache network. The
work in \cite{rosensweig2010approximate} provided an approximate model
with LRU policy. In \cite{psaras2011modelling}, the authors first
proposed a Markov chain based mathematical model for a single router
that models the time proportion that a content is cached, and then
extended to a class of cache networks. 
In \cite{gallo2012performance}, the authors focused on the cache hit-ratio
estimation with random replacement policy whose performance is
shown to be similar to that of LRU, meaning that random replacement can
be a low-cost candidate cache policy for the caches with high access
rates.  The authors in \cite{CSsize} compare the performance of a cache
network for different cache sizing policies, depending on the importance
of a cache node, e.g., degree centrality. Due to technical challenges in
analyzing general topologies, a special class of topologies such as tree or cascade
has been popularly studied \cite{muscariello2011bandwidth,
  carofiglio2011modeling}.   
  
The asymptotic analysis of cache networks has been studied only in
wireless (multi-hop) networks, to the best of our knowledge.  In
\cite{gitzenis2012asymptotic}, it was proved the required link capacity decreases from $O(\sqrt{n})$ to down to $O(1)$ using
caches, where $n$ is the number of nodes. This is
due to the reduction of wireless interference induced by the decrease in
necessary transmissions in presence of caching. In
\cite{azimdoost2012throughput},  a dynamic content change at
caches was modeled by abstracting cache dynamics with limited lifetime
of cached content. They showed that
maximum throughput becomes $1/\sqrt{n}$ and
$1/\log n$ for grid and random networks, compared to $1/n$ and
$1/\sqrt{n\log n}$ for non-cache network.

\section{System Model and Problem Statement}\label{sec:model}

\subsection{Model}

\noindent{\bf \em Network and content servers.} 
We consider a sequence of graphs $\mathcal G_n=(\mathcal V_n,\mathcal
E_n)$, where $\mathcal V_n$ is the set of nodes or caches with
$|\mathcal V_n|=n$ and $\mathcal E_n\subset \mathcal V_n\times\mathcal
V_n$ describes neighboring relations between caches. In addition, we let
$\mathcal C_n$ be the set of contents, and $\mathcal S_n$ be the set of
servers containing the original contents. For notational simplicity, we
will drop the subscript $n$ for all quantities that depends on $n$,
unless confusion arises.  We assume that contents are of equal size and
each content $c \in\mathcal C$ is stored in a single server, say $s_c$,
and each server $s_c\in \mathcal S$ is attached to a randomly chosen
node $v_c := v_{s_c} \in \mathcal V.$ In this paper, we consider the
case of one server per one content, but we remark that our results can
be easily extended to multiple-server
cases. 

\smallskip
\noindent{\bf \em Content requests and routing.}
We assume that there are exogeneous requests for contents at each cache,
and locations of servers and content requesting nodes are uniformly at
random in $\mathcal V$.  In this environment, we assume that network capacity is
large enough to ignore the negligible values
such as waiting time at caches, and only consider the service time, denoted as the
delay (see its formation definition in Section~\ref{sec:metric}).
Additionally, we assume each request is independent with others and the
request rate of content $c_i \in \mathcal C=\{c_1, c_2, \ldots\}$
is proportional to a Zipf-like distribution with parameter $\alpha>0$:
\begin{equation}
p_i=\frac{K}{i^{\alpha}},\label{eq:zipf}
\end{equation}
where the normalizing constant $K$ is such that 
$\frac{1}{K}=\sum_{i=1}^{\nC} 1/i^{\alpha}.$ For a higher value
$\alpha,$ we sometimes say that a cache network is with {\em higher
popularity bias}, i.e., content popularity difference is high in that
network.  When a request for content $c\in \mathcal C$ arrives at cache
$v$, it is forwarded along the path given by some routing algorithm,
e.g., the shortest path routing in $\mathcal G$, from the cache $v$ to
the server $v_c$.  
The request generates HIT at cache $v$ if
$c$ is located at $v$ on the routing path or MISS otherwise. In the MISS
event, the request keeps being forwarded to a next cache in the routing
path until reaching
to the server $v_c.$

\smallskip

\noindent{\bf \em Caches and policies.}
Each cache $v\in\mathcal V$ stores a set of contents $\mathcal C_v\subset \mathcal C$ independently, and has the cache size $b_v:=|\mathcal C_v|\geq 0$ with the network-wide cache budget $B = \sum_{v \in \mathcal V} b_v$. The primary goal of this paper is to choose appropriate $[b_v]_{v \in \mathcal V}$ and $[\mathcal C_v]_{v \in \mathcal V}$ for the high performance of the cache networked system. Clearly, how many and which contents can be stored in each cache is governed by  {\em content placements} and {\em cache sizing policies}. 
For content placements, the following rules are studied in this paper:
 \smallskip
  
 \begin{compactitem}
 \item 
{\bf \em URP (Uniformly Random Policy).}  Each cache
 contains $b_v$ contents which are chosen from $\mathcal C$ uniformly at
 random.

 \item
{\bf \em PPP (Pure Popularity-based Policy).}  Each
 cache contains $b_v$ contents following the `pure' (or `exact') content
 popularity distribution, i.e., for a popularity parameter $\alpha.$

 \item
{\bf \em TPP (Tilted Popularity-based Policy).}  We
  also study more generalized popularity-based policies: each contains
  $b_v$ contents following the Zipf-like distribution
  with parameter $\beta$ (which may not be equal to $\alpha$).  Since we
  found that the choice $\beta=\alpha/2$ is optimal in some sense over
  many scenarios, we only focus on such a choice (see
  Section~\ref{sec:main_homo} for details).

\item 
{\bf \em TPP-C (TPP with Cutting).}  This policy is a
  variant of TPP, with main difference that contents under some
  threshold in the popularity ranking are not cached.  Specifically,
  under the assumption that the average routing distance $\bar{d}$ is
  given, we compute the content index $\hat{i} = \min \{ s \cdot \bar{d}
  , \nC \}$ (cf, $s$ is the per-node cache size) and only the contents
  in $\{c_1, c_2, \ldots, c_{\hat{i}}\} \subset \mathcal C$ are randomly
  cached, following the Zipf-like distribution with $\beta=\alpha/2$.

 \end{compactitem}

\smallskip
For cache sizing policies, we separately study the following:

\smallskip
\begin{compactitem}
\item 
{\bf \em Homogenous per-node cache size.}  All caches
 have the same size such that $b_v=\frac{B}{n}$ for all $v \in \mathcal
 V$ (see Section \ref{sec:equal_size}).
 
\item 
{\bf \em Heterogeneous per-node cache size.}  We also consider
 a setting concentrating cache budgets on more influential nodes, for
 instance,   nodes with high `betweenness centrality' that quantifies the fraction of
 shortest paths that pass through a node, i.e., cache budgets
 $[b_v]$ are different among nodes (see Section \ref{sec:non_equal_size}). 
\end{compactitem}

\subsection{Performance Metric}
\label{sec:metric}

In this section, we introduce the performance metric of our interest for cache networks.
We define the delay of a content request as the number of (expected) hops until it finds the desired content, i.e.,
HIT occurs.  Formally, let random variable $X_i$  be the delay of the $i$-th request for some content in
the (entire) cache network.  Then, the asymptotic average
delay $\Delta$ of the cache network is defined as follows:

$$\Delta \triangleq \lim_{N\to \infty} \bbexpect{
  \frac1{N}\sum_{i=1}^{N} X_i}$$ 
where it is not hard to
check that the limit always exists given system setups $\mathcal G, [\mathcal C_v], \alpha,$ and a routing policy.

\begin{table*}[t!]
  \centering
  \renewcommand{\tabcolsep}{0.4mm}
  \renewcommand{\arraystretch}{1.8} 
  \caption{Homogeneous cache size: Delay of five static content
    placement policies. TPP-C-AVG corresponds to TPP-C with the average
    routing distance (and thus delay upper-bound from Jensen's
    inequality) \label{table:total} }
  \vspace{-0.2cm}
  {\small
  \begin{tabular}{|c||c|c|c|c|c||c|} \hline 
    & URP&PPP &LBND  & \specialcell{TPP and \\ TPP-C ($s \cdot \bar{d} \ge \nC
    $)}&TPP-C ($s \cdot \bar{d} < \nC $)& TPP-C-AVG \\ \hline \hline

    $2<\alpha$& $\Theta\Big(\min [ d,\frac{\nC}{s} ]\Big)$&$O\Big(\min[
    \frac{(ds)^{1/\alpha}}{s}, \frac{\nC}{s} ] \Big)$& $\Theta(1)$ &
    $\Theta(1)$ & $O\Big(\frac{d}{(s\bar{d})^{\alpha-1}}\Big)$&$\Theta (1)$\\ \hline 

    $\alpha=2$& $\Theta\Big(\min [ d,\frac{\nC}{s} ]\Big)$ &$O\Big(\min
    [\sqrt{\frac{d}{s}}, \frac{\nC}{s}]\Big)$& $\Theta\Big(\frac{\log (\min
      [s\cdot d , \nC ])}{s}\Big)$& 	 \specialcell{$O \Big(\min [d,
      \frac{\log^2(\nC)}{s},$ \\ $\frac{\log(\nC)
  \log(s\cdot d)}{s} ] \Big)$} &$O\Big(\frac{1}{s} \max [\log^2\bar{d}, \frac{d}{\bar{d}}]\Big)$ & $ O\Big(\frac{\log^2 (\min
  [s\cdot \bar{d} , \nC ])}{s}\Big)$\\ \hline

    \specialcell{$1< \alpha$\\$<2$}& $\Theta\Big(\min [ d,\frac{\nC}{s} ]\Big)$ &$O\Big(\min [
    \frac{(ds)^{1/\alpha}}{s}, \frac{\nC}{s} ]\Big)$& $\Theta\Big(\frac{(\min
      [s\cdot d, \nC ])^{2-\alpha}}{s}\Big)$& \specialcell{$O\Big ( \min [ d,
      \frac{\nC^{2-\alpha}}{s},$\\ $\frac{\nC^{(2-\alpha)\frac{\alpha-1}{\alpha}}d^{\frac{2}{\alpha}-1}}{s^{2-2/\alpha}}  ] \Big )$}&$O\Big((s\bar{d})^{1-\alpha} \max [\bar{d},d ]\Big)$ & $O\Big(\frac{(\min
  [s\cdot \bar{d}, \nC ])^{2-\alpha}}{s}\Big)$\\ \hline 

$\alpha=1$& $\Theta\Big(\min [ d,\frac{\nC}{s} ]\Big)$&$\Theta\Big(\min [
d,\frac{\nC}{s} ]\Big)$& $\Theta\Big(\min [ d, \frac{\nC}{s \cdot \log \nC } ] \Big)$ &  $O \Big( \min [ d,  \frac{\nC}{s \cdot \log \nC}   ]   \Big )$&$O\Big(\max [ \frac{\bar{d}}{\log \nC}, d] \Big)$& $O \Big( \min [ \bar{d},  \frac{\nC}{s \cdot \log \nC}   ]  \Big)$\\ \hline

    \specialcell{$0 < \alpha$\\$ <1$}& $\Theta\Big(\min [ d,\frac{\nC}{s} ]\Big)$&$\Theta\Big(\min [ d,\frac{\nC}{s} ]\Big)$& $\Theta\Big(\min [ d,\frac{\nC}{s} ]\Big)$ & $O \Big( \min [ d,  \frac{\nC}{s}   ]  \Big)$&$O \Big(\max [ \bar d\cdot (\frac{s\cdot \bar d}{\nC})^{1-\alpha}, d ] \Big)$ & $O \Big( \min \{ \bar{d},  \frac{\nC}{s}   ]  \Big)$\\ \hline
  \end{tabular}
}
\vspace{-0.4cm}
\end{table*}

For a fixed $d >0,$ let $\Delta(d)$ be the ``expected'' (or average) delay when the routing
distance between a content requesting node and the server is $d,$ where 
the expectation is taken with respect to the randomness in the requested
content, content placement policy and cache sizing policy. 
More formally, for a given distance $d,$
\begin{eqnarray}
  \label{eq:expected_delay}
\Delta (d) = \sum_{i=1}^{\nC} p_i  \xi_i (d),
\end{eqnarray}
where $\xi_i (d)$ denotes the expected delay of contents $c_i$ for
a given distance $d$ 
under a (fixed) routing policy.  
However, note that $d$ is a also random variable, when a randomly chosen
content requesting node is assumped. Thus, the actual average delay
$\Delta$ is given by, from \eqref{eq:expected_delay},
\begin{eqnarray}
  \label{eq:avg_delay}
  \Delta = \expect{\Delta(d))} = \sum_{d} f_d \Delta(d) = \sum_{d} f_d \sum_{i=1}^{\nC} p_i  \xi_i (d),\label{eq:average_delay}
\end{eqnarray}
where the expectation is taken over the distribution of random variable $d$ and
$f_d$ is its probability which
relies on the underlying topology of $\mathcal{G}$ and a given routing policy. 

Our objective is to study the asymptotic order of $\Delta$ under various
setups, by studying $\Delta(d),$ which are analyzed in Sections~\ref{sec:equal_size}
and \ref{sec:non_equal_size}. 
This study will asymptotically quantify the fundamental
performance gains generated by the network of caches, which is expected
to give practical implications into how we should design a cache
network.

\section{Homogenous Per-node  Cache Size}
\label{sec:equal_size}

\subsection{Approach and LBND policy}\label{sec:approaching}
In this section, we first focus on the case when each node has equal
cache budget $s = s_v = B/n$ for all caches $v.$ Content placement
policies considered in our paper are mostly all identical random ones and do not differentiate
particular caches. Hence, 
for a given routing path with distance $d,$ the average delay $\xi_i (d)$ for
content $i$ depends only on the distance $d$ and the cache
hit probability $h_{c_i}$ of content $i$ at any arbitrary node, which is
simply given by:
\begin{eqnarray}
  \label{eq:avg_delay_content}
\xi_i(d) =  \sum_{l=1}^{d-1} l \cdot h_{c_i} \cdot (1-h_{c_i})^{l-1} +
d \cdot (1-h_{c_i})^{d-1}.
\end{eqnarray}

In addition to four content placement policies introduced in Section
\ref{sec:model}, we also consider an unrealistic ideal policy, which we call
{\bf LBND} (Lower BouND), that provides delay lower bounds on $\Delta(d),$ i.e., any
policy cannot beat it. In LBND, for any routing path between a content requesting node and the server, contents are assumed to be placed on caches with descending order
of popularity from the most polular contents such as $\{c_1, \dots , c_s
\},$ $\{c_{s+1}, \dots , c_{2s} \}, \ldots.$ 
Clearly, this is unrealistic because such a
popularity-based descending ordering for {\em any} requesting node and
server is impossible.  Note that in LBND, $\xi_i (d),$ the average
delay of content $c_i$ for given distance $d,$ is $\min\{ \lceil i/s
\rceil , d\}$ where $\lceil x \rceil$ indicates the minimum integer satisfying $\lceil x \rceil \ge x.$

\subsection{Main Result}\label{sec:main_homo}
\begin{theorem}[Delay for homogeneous cache size]
  \label{thm:homo}
  For a given routing distance $d$ and the average routing distance
  $\bar{d}$ between an arbitrary pair of content
  requester and content server, the average delay $\Delta(d)$ scales as
  those in Table~\ref{table:total} under homogenous per-node cache
  size. 
\end{theorem}

\smallskip
The proof of Theorem~\ref{thm:homo} is in Section~\ref{sec:proof-theor-refthm:h}. 

Here, we first provide interpretations of Theorem~\ref{thm:homo}. For ease of
explanation, we assume that $s = s_n = O(1),$ which is the most
interesting case, because our natural interest lies in whether there is
a delay reduction via a small amount of cache budget. For a constant per-node
cache size, the results in Table~\ref{table:total} can be conveniently 
explained by diving the regimes into (i) $C \ll d$ and $C \gg d$
(in the asymptotic sense). 

\smallskip
\begin{compactenum}[\em (a)]
\item We provide upper bounds (i.e., $O(\cdot)$) on $\Delta(d)$ for
PPP, TPP, and TPP-C, and upper/lower bounds
  (i.e., $\Theta(\cdot)$) on $\Delta(d)$ for URP and
  LBND policies.

\item As expected, the caching gains of popularity-based policies such
  as PPP, TPP, TPP-C increase as content
  popularity bias parameter $\alpha$ grows.

\item For $\alpha >2$ (very high popularity bias), TPP and TPP-C are
  order-optimal.

\item In case of $\nC \ll d,$ TPP and TPP-C outperforms
  PPP, where even PPP's delay becomes  just the same as that
  of URP, and TPP/TPP-C is very close to even LBND.
  This is because when $\nC \ll d,$ there is a large number of caching places
  from the requester to the corresponding server, in which case caching
  less popular contents such as TPP/TPP-C significantly helps in
  reducing delay, whereas a policy giving too much priority to more
  popular contents such as PPP is not highly effective. 

\item However, in case of $\nC \gg d,$ the opposite occurs, i.e., due to
  lack of caches in the routing path, to reduce delay, more popular
  contents should be cached with high probability. Thus, PPP outperforms TPP, where TPP is no
  better than URP. 

\item TPP-C can be regarded as an {\em adaptive} policy that works well for both cases,
  because it tends to cache more kinds of caches when $\nC \ll d,$ and
  focus on more popular contents when $\nC \gg d,$ by adaptively determining
  the contents that should not be cached. 

\item As presented in \eqref{eq:avg_delay}, our analytical result
  $\Delta(d)$ in Theorem~\ref{thm:homo} can be plugged into the equation
  $\Delta = \sum_{d} f_d \Delta(d)$ to obtain the final average delay,
  once the distribution of routing distance $f_d$ is known. However, in
  case when only {\em average} routing distance is available, our result
  is of great use, because from Jensen's inequality and concavity of $\Delta$,
  \begin{eqnarray}
    \label{eq:jensen}
    \Delta = \expect{\Delta(d)} \leq \Delta(\expect{d}),
  \end{eqnarray}
and by replacing $d$ in Table~\ref{table:total} by the average routing distance $\bar{d} =
\expect{d},$ at least delay upper-bounds can be computed. In fact, we
present this for TPP-C, named TPP-C-AVG in Table~\ref{table:casestudy}, which shows delay performance
being very close to LBND. 

\end{compactenum}

\medskip
\noindent{\bf \em Why $\alpha/2$ in TPP and TPP-C?}
As mentioned earlier, TPP is the policy that provides more chances for
less popular contents to be cached than PPP, and TPP-C is based on TPP
with cutting the contents with ``very low'' popularity.  But, why
$\alpha/2$ in TPP/TPP-C, rather than $\alpha/3$ or $\alpha/10$?  Just for
simplicity of exposition, assume $s=1,$ i.e., each node can cache only
one content, and also assume that the routing distance $d$ is extremely
large, just like the regime $d \gg C.$ Now consider a cache placement
policy under which content $c_i$ is cached in each cache with
probability $q_i.$ Note that a special case when $q_i = p_i$ corresponds
to PPP (because PPP directly applies the content popularity distribution
to the cache placement distribution) Then, the expected delay
$\Delta(d)$ becomes:
\begin{equation*}
  \Delta (d)  =  \sum_{i=1}^{\nC} p_i \cdot \frac{1}{q_i} =\big(\sum_{i=1}^{\nC}p_i \frac{1}{q_i}\big) \big(\sum_{i=1}^{\nC}{q_i}\big) \ge \big(\sum_{i=1}^{\nC}{p_i}^\frac{1}{2}\big)^2,
\end{equation*}
where the last inequality comes from the Cauchy-Schwarz
inequality. In Cauchy-Schwarz inequality, it is widely known that the equality holds if and only if there is
some constant $k$ such that  $p_i \frac{1}{q_i}=k \cdot q_i$ for all
$i.$ Therefore, $\Delta (d)$ is minimized when 
$q_i  \propto i^{-\frac{\alpha}2}, $
and the minimum value is
$\big(\sum_{i=1}^{\nC}{p_i}^\frac{1}{2}\big)^2.$ This is why $\alpha/2$
is selected for TPP/TPP-C.

\subsection{Application to Power-law and Erd\"os$-$R\'enyi graphs}
As case studies, we now apply Theorem \ref{thm:homo} to popular random
graphs: Power-law (PL) and Erd\"os$-$R\'enyi (ER) graphs, where we
assume a shortest-path based request routing algorithm, and $s = \Theta(1)$
and $|\set{C}| = \Theta(n).$

\begin{table}[t!]
  \centering
  \tabcolsep=0.7mm
  \renewcommand{\arraystretch}{1.6} 
  \caption{Orders of delay with URP, PPP, TPP, TPP-C, and LBND policies for average distance $\bar{d}$ in case study  \label{table:casestudy} }
  \vspace{-0.15cm}
  {\small
  \begin{tabular}{|m{1.5cm}||m{0.8cm}|m{1.4cm}|m{0.8cm}|m{1.4cm}|m{1.4cm}|} \hline 
   & URP&PPP & TPP&TPP-C&LBND  \\ \hline 
 $2<\alpha$& $O(\bar{d})$&$O((\bar{d})^{1/\alpha})$& $\Theta(1)$ &$\Theta(1)$ & $\Theta(1)$\\ \hline 
$\alpha=2$& $O(\bar{d})$ &$O(\sqrt{\bar{d}})$& $O(\bar{d})$& $O ( \log^2(\bar{d}))$& $O( \log\bar{d})$\\ \hline 
 $1< \alpha<2$& $O(\bar{d})$ &$O((\bar{d})^{1/\alpha})$& $O(\bar{d})$&$O ((\bar{d})^{2-\alpha})$& $O((\bar{d})^{2-\alpha})$ \\ \hline 
$0<\alpha\le 1$& $O(\bar{d})$ &$O(\bar{d})$& $O(\bar{d})$&$O(\bar{d})$& $O(\bar{d})$ \\ \hline
  \end{tabular}
}
\end{table}


In the PL graph, the fraction of nodes with
degree $i$ is proportional to $1/i^\gamma$ for some
constant $\gamma>0.$ If the average degree is strictly greater
than 1, and $2<\gamma<3$, it is known that 
the average routing distance under the shortest path routing
is $\bar d=\Theta(\log n/ \log \log n)$ \cite{chung2002average}. 
The ER-graph is constructed by randomly connecting two
nodes with some probability, say $p.$ 
If $np$ is of order $\log n$, then the graph
almost surely contains a giant component of size of order $n$
connected with high probability, and it is known in \cite{draief2010epidemics} that
the average routing distance under the shortest path routing is $\bar d=\Theta\left(\frac{\log n}{\log np}\right).$ 
Using those facts about the average routing distances under two example
random graphs and applying $\bar{d}$ for upper-bounds from Jensen's
inequality (as in TPP-C-AVG), we obtain the delay orders for various
content placement policies, shown in Table~\ref{table:casestudy},
where major interpretations are summarized
as:  In both graphs,  
\begin{itemize}
\item For $0<\alpha<1$, all policies have $O(\bar{d})$ delay, i.e., 
  no cashing gain occurs from content popularity.
\item For $\alpha>2$,  TPP and TPP-C have $\Theta(1)$ delay (i.e., order-optimal).
\item For any $\alpha>0$, TPP-C policy outperforms other policies. 
\end{itemize}

\subsection{Proof of Theorem~\ref{thm:homo}}\label{sec:proof-theor-refthm:h}

\noindent{\bf \em Proof for URP.}
We first derive the hit probability $h_{c_i}.$ Since each cache have $s$
rooms for the content placement and there are $\nC$ contents,
$\binom{\nC}{s}$ possible contents configurations are located over all
caches uniformlyㅅ at random. Moreover, since the number of
configurations including content $c_i$ is $\binom{\nC-1}{s-1},$ the hit
probability of content $c_i,$ $h_{c_i}$ is $\frac{\binom{\nC-1}{s-1}}{\binom{\nC}{s}}=\frac{s}{\nC},$
 which is the
same over all contents. 
Hence, one can easily compute $\Delta(d)=\xi_i(d)$ as follows:
\begin{multline*}
\Delta(d)=\xi_i(d) =  \sum_{l=1}^{d-1} l \cdot h_{c_i} \cdot (1-h_{c_i})^{l-1} +
d \cdot (1-h_{c_i})^{d-1}
  \cr =  \frac{1-(1-h_{c_i})^{d}}{h_{c_i}} = \frac{\nC}{s}\left(1-\left(1-\frac{s}{\nC}\right)^{d}\right)\cr
=\frac{\nC}{s}\left(1-\left(\left(1-\frac{s}{\nC}\right)^{\frac{\nC}{s}}\right)^{\frac{d s}{\nC}}\right)
=\Theta\Big(\frac{\nC}{s}\left(1-e^{-\frac{d s}{\nC}}\right)\Big)
\end{multline*}
First, when $d\leq\frac{\nC}{s}$,  one can check that
\begin{align*}
\Delta(d)&=\Theta\left(\frac{\nC}{s}\left(1-e^{-\frac{d s}{\nC}}\right)\right)\cr
&=\Theta\left(\frac{\nC}{s}\left(1-\left(1-\frac{d s}{\nC}\right)\right)\right) =\Theta( d).
\end{align*}
On the other hand, when $d\geq\frac{\nC}{s}$,  it follows that
$\Delta(d)=\Theta(\frac{\nC}{s}\left(1-e^{-\frac{d s}{\nC}}\right))
=\Theta(\frac{\nC}{s}).$
This completes the proof. 

\smallskip
\noindent{\bf \em Proof for PPP.}
To begin with, one can lower bound the hit probability under PPP by:
$h_{c_i} \ge 1 - (1-p_i)^{s}.$ Then, from the equation
(\ref{eq:avg_delay_content}), we have
\begin{align}
\xi_i(d)& = \frac{1-(1-h_{c_i})^{d}}{h_{c_i}}  ~ \le \frac{1}{h_{c_i}} ~= 1+\frac{(1-{p_i})^s}{1 - (1-{p_i})^s}
\cr &= 1 + \frac{(1-{p_i})^s}{p_i \cdot \sum_{k=0}^{s-1}
   (1-{p_i})^k}  ~ \le 1 + \frac{1}{s \cdot p_i}. \label{eq:ppp}
\end{align}
Moreover, we know $\xi_i(d) \le d.$ 
Therefore, using these observations, for given routing distance $d$,
the expected delay becomes
\begin{align}
\Delta(d) &\le 1+\sum_{i=1}^{\nC} p_i \cdot \min \left\{\frac{1}{s \cdot p_i} ,d\right\}
\cr &\le 1+ \sum_{i=1}^{i^*} p_i \cdot \frac{1}{s \cdot p_i}  + \sum_{i=i^*+1}^{\nC}p_i \cdot  d, \label{eq:epp}
\end{align}
where $i^*$ is an integer which we will decide later.
Now, we will compute \eqref{eq:epp} depending on ranges of $\alpha .$

\noindent\underline{(i) $\alpha>1$:} 
In this case, $K = \Theta(1)$ which is the normalizing constant for the popularity distribution.
Thus, when we set
$i^* = \Theta ( \min\{ \nC, (s\cdot d)^{1/\alpha} \} )$ and obtain
\begin{multline*}
\Delta(d) \le 1+ \sum_{i=1}^{i^*} p_i \cdot \frac{1}{s \cdot p_i}  +
\sum_{i=i^*+1}^{\nC}p_i \cdot  d 
\end{multline*}
\begin{multline*}
 \le  1+ \frac{i^*}{s} + O( \frac{d}{(i^*)^{1-\alpha}} ) 
 =  O\left(\frac{\min\{ \nC, (s\cdot d)^{1/\alpha} \}}{s} \right). 
\end{multline*}

\noindent\underline{(ii) $\alpha = 1$:} 
In this case, $K = \Theta(\frac{1}{\log (\nC)})$ and we set
$i^* = \Theta ( \min\{ \nC, s\cdot d \} ).$
Then, we have
\sqeq
\begin{multline*}
\Delta(d)  \le  1+ \frac{i^*}{s} + O\left( d-\frac{\log (i^*)}{\log (\nC)}d \right) 
  =  O\left( \min\left\{ \frac{\nC}{s}, d \right\} \right). 
\end{multline*}
\unsqeq

\noindent\underline{(iii) $0<\alpha < 1$:} 
In this case, $K = \Theta((\nC)^{\alpha-1})$ and we set 
$i^* = \Theta ( \min\{ \nC, s\cdot d \} ).$
Then, we have
\begin{align*}
\Delta(d) & \le  1+ \frac{i^*}{s} + O\left( \frac{(\nC)^{1-\alpha}-(i^*)^{1-\alpha}}{(\nC)^{1-\alpha}}d \right) \cr
 & =  O\left( \min\left\{ \frac{\nC}{s}, d \right\}\right ). 
\end{align*}
Therefore, Theorem~\ref{thm:homo} for PPP follows.

\smallskip
\noindent{\bf \em Proof for TPP.}
By using the similar arguments for deriving (\ref{eq:epp}) (using $q_i$ instead of $p_i$), 
one can have that the
expected delay $\Delta (d)$ is bounded as follow :
\begin{align} 
\Delta (d)&\le 1+ \sum_{i=1}^{i^*} p_i \cdot \frac{1}{s \cdot q_i} +
\sum_{i=i^*+1}^{\nC}p_i \cdot  d \cr
 & =   1+ \sum_{i=1}^{i^*} \frac{K}{K'} \cdot \frac{1}{s \cdot
   i^{\alpha / 2}} +
\sum_{i=i^*+1}^{\nC}\frac{K}{i^{\alpha}} \cdot  d, \label{eq:etp}\end{align}
where $i^*$ is an integer which we will decide later and $K'$ is the normalizing constant for the probability distribution
corresponding to $q_i$. Now we will do the case study depending on $\alpha$ similarly as we did before.

\noindent\underline{(i) $\alpha > 2 $:} 
In this case, $K=\Theta(1)$ and $K'=\Theta(1).$ 
Therefore, $\Delta (d) = O (1) $ where we set $i^* = 
\nC$ in (\ref{eq:etp}).

\noindent\underline{(ii) $\alpha = 2$:}
In this case, $K=\Theta(1)$ and $K'=\Theta(\frac{1}{\log (\nC)}).$ Thus, (\ref{eq:etp}) becomes
\begin{align*}
\Delta (d) &=  1+O \left(  \log (\nC)\cdot \frac{\log (i^* )}{s 
   } \right) + O \left( \frac{d}{i^*} \right)\\
&=O \left(\min \left\{d, \frac{\log^2(\nC)}{s}, \frac{\log(\nC)
  \log(s\cdot d)}{s} \right\}\right ),
\end{align*}
where we choose $i^* = \max \left\{ 1, \min \left\{\nC , \frac{s\cdot d}{\log (\nC)} \right\} \right\} $.

\noindent\underline{(iii) $1<\alpha<2 $:}
In this case, $K=\Theta(1)$ and
$K'=\Theta(\frac{1}{(\nC)^{1-\alpha/2}}).$ Thus, (\ref{eq:etp}) becomes
\begin{align*}
\Delta(d)&= 1+ O \left((\nC)^{1-\frac{\alpha}{2}}\cdot
  \frac{(i^*)^{1-\frac{\alpha}{2}}}{s} \right) + \left( (i^*)^{1-\alpha} \cdot d \right)\\
&=O \left( \min \{ d,
  \frac{\nC^{2-\alpha}}{s},\frac{\nC^{(2-\alpha)\frac{\alpha-1}{\alpha}}d^{1-\frac{2}{\alpha}}}{s^{2-2/\alpha}}
  \}  \right),
\end{align*}
where we choose $i^* = \max\left\{ 1, \min \left\{\nC , \frac{(s\cdot
  d)^{2/\alpha}}{(\nC)^{2/\alpha -1}} \right\}  \right\}.$ 

\noindent\underline{(iv) $\alpha = 1 $:}
In this case, $K=\Theta(\frac{1}{\log \nC})$ and
$K'=\Theta(\frac{1}{(\nC)^{1-\alpha/2}}).$ Thus, (\ref{eq:etp}) becomes
\begin{align*}
\Delta(d)&= O \left(\frac{\nC^{\frac{1}{2}}}{\log \nC}\cdot
  \frac{(i^*)^{\frac{1}{2}}}{s} + \frac{1}{\log \nC} \log \frac{\nC}{i^*} \cdot d \right)\\
&=O \left( \min \left\{ d,
  \frac{\nC}{s \cdot \log \nC} 
  \right\}  \right),
\end{align*}
where we choose $i^* = \max \{ 1, \min \{\nC , \frac{(s\cdot
  d)^{2}}{\nC} \}  \}.$ 

\noindent\underline{(v) $0<\alpha<1 $:}
In this case, $K=\Theta(\frac{1}{\nC^{1-\alpha}})$ and
$K'=\Theta(\frac{1}{(\nC)^{1-\alpha/2}}).$ Thus, (\ref{eq:etp}) becomes
\begin{align*}
\Delta(d)&= O \left(\nC^{\frac{\alpha}{2}}\cdot
  \frac{(i^*)^{1-\frac{\alpha}{2}}}{s} + \frac{\nC^{1-\alpha} - (i^*)^{1-\alpha}}{\nC^{1-\alpha}} \cdot d \right)\\
&= O \left( \min \left\{ d,
  \frac{\nC}{s}  \right\}  \right),
\end{align*}
where we choose $i^* = \Theta (\min \{\nC , s\cdot d \} ).$
Therefore, Theorem~\ref{thm:homo} for TPP follows.

\smallskip
\noindent{\bf \em Proof for TPP-C.}
TPP-C is the same policy with TPP when $\nC \le s
\cdot \bar{d}$. Thus, in this proof, we will only consider the case $\nC > s \cdot \bar{d} .$ 
Then, similarly as we did before, we have
\begin{align} 
\Delta (d) &\le 1+ \sum_{i=1}^{\hat{i}} p_i \cdot \frac{1}{s \cdot \hat{q_i}} + \sum_{i=\hat{i}+1}^{\nC}p_i \cdot  d \cr
 & =   1+ \frac{K}{s \cdot M^2} + \sum_{i=\hat{i}+1}^{\nC}\frac{K}{i^{\alpha}} \cdot  d, \label{eq:metp}\end{align}
where ${M} := \left(\sum_{i=1}^{\hat{i}}\frac{1}{i^{\alpha/2}} \right)^{-1}$ and we set $\hat{i}= s\cdot \bar d $. 

\noindent\underline{(i) $2<\alpha $:} In this case, $M=\Theta(1)$
and $  1+ \frac{2K}{s \cdot M^2}  = \Theta (1).$ 
Thus, we have
\begin{align*}
\Delta(d)=O\left(\sum_{i=\hat{i}+1}^{\nC}\frac{K}{i^{\alpha}} \cdot  d\right) & =  \Theta
\left(d\cdot ( \hat{i}^{1-\alpha} - \nC ^{1-\alpha}  )  \right)\\
& =  \Theta \left(d((s\cdot \bar{d} )^{1-\alpha} - \nC ^{1-\alpha}
  ) \right).
\end{align*}

\noindent\underline{(ii) $\alpha = 2$:}
In this case $\frac{1}{M}=\Theta(\log (\hat{i}))$ and (\ref{eq:metp}) becomes
\begin{align*} 
\Delta (d) &=O \left(\frac{1}{s}\log^2 (s\cdot \bar{d}) + d\cdot
  \left(\frac{1}{s \cdot \bar{d}} - \frac{1}{\nC}\right)\right).
\end{align*}

\noindent\underline{(iii) $1<\alpha<2 $:}
In this case $\frac{1}{M}=\Theta((\hat{i})^{1-\alpha/2})$ 
and $K=\Theta(1)$. Thus, (\ref{eq:metp}) becomes,
\begin{align*} 
\Delta(d) &= O \left(\frac{(s \cdot \bar{d})^{2-\alpha}}{s}+d\cdot(
  {(s\cdot \bar{d})}^{1-\alpha} - {\nC}^{1-\alpha} )\right).
\end{align*}

\noindent\underline{(iv) $\alpha = 1 $:} 
In this case, $K = \Theta (1/\log \nC)$ and (\ref{eq:metp}) becomes
\begin{align*} 
\Delta(d) &=O \left(\frac{\bar{d}}{\log \nC}+d\cdot(1- \frac{\log ( s\cdot \bar{d})}{\log \nC})\right).
\end{align*}

\noindent\underline{(v) $0<\alpha<1 $:}
In this case, $K = \Theta (\nC ^{\alpha - 1})$ and (\ref{eq:metp}) becomes
\begin{align*}
\Delta(d) &= O \left(\frac{s\cdot \bar d}{s}(\frac{s\cdot \bar
    d}{\nC})^{1-\alpha}+d \cdot (1-(\frac{s\cdot \bar d}{\nC})^{1-\alpha})\right) \cr
&=O \left( \bar d\cdot (\frac{s\cdot \bar d}{\nC})^{1-\alpha} + d
  \cdot (1-(\frac{s\cdot \bar d}{\nC})^{1-\alpha}) \right).
\end{align*}
Therefore, Theorem~\ref{thm:homo} for  TPP-C follows.

\smallskip
\noindent{\bf \em Proof for LBND.}
Since $\xi_i (d)=\min\{ \lceil i/s \rceil , d\}$ under LBND, it follows that 
\begin{multline}
\Delta(d) =  \sum^{\nC}_{i=1} p_i \cdot \xi_i (d)
 =  \sum^{s}_{i=1} p_i \cdot \min \{d, 1\}+ \sum^{2s}_{i=s+1} p_i
 \cdot \min \{d, 2\}\cr
 +
   \cdots + \sum^{\nC}_{i=(\lceil \nC/s \rceil -1)s+1} p_i \cdot
\min \left\{d,\left\lceil \frac{\nC}{s} \right\rceil \right\} \label{eq:2}
\end{multline}
Let $z=\min \left\{d, \left\lceil \frac{\nC}{s} \right\rceil \right\}$ and
we can rewrite \eqref{eq:2} as follows: $\Delta(d) = \sum_{l=1}^{z}\sum^{\nC}_{i=1+(l-1)s}
K/i^{\alpha}.$

We now prove the desired results for different values of $\alpha.$ To
this end, we first find the order of $K$ and the order of
remaining terms, separately,  and then, we will combine
the results to get the delay order. 


\noindent\underline{(i) $2<\alpha$:} In this case,
$K=\Theta(1)$ since
\begin{align}
\frac{1}{K} =\sum^{\nC}_{i=1}\frac{1}{i^{\alpha}} = \Theta \left( \int_{x=1}^{\nC}
      \frac{1}{x^\alpha}   \text{d}x \right) = \Theta(1). \label{eq:k}
\end{align}
In addition, we compute that

\begin{align*} 
\sum_{l=1}^{z}\sum^{\nC}_{i=1+(l-1)s}
\frac{1}{i^{\alpha}} &=  \Theta \left(1+\int_0^{z-1}\int_{k\cdot
    s+1}^{\nC}\frac{1}{x^{\alpha}}\text{d}x \text{d}k \right)
\cr &=  \Theta (1+\int_0^{z-1} ({k\cdot s +1})^{1-\alpha}- \nC^{1-\alpha} \text{d}k)
\cr &=  \Theta \left(1 + \frac{1-(zs)^{2-\alpha}}{s}\right)= \Theta (1). 
\end{align*}
Thus, the delay order becomes $\Theta (1).$

\noindent\underline{(ii) $\alpha = 2$:} In this case, as \eqref{eq:k}, 
$K = \Theta (1) $ and 
\begin{multline*} 
\sum_{l=1}^{z}\sum^{\nC}_{i=1+(l-1)s}
\frac{1}{i^{2}} =  \Theta \left(\int_0^{z-1}\int_{k\cdot
    s+1}^{\nC}\frac{1}{x^{2}}\text{d}x \text{d}k \right)
\cr =  \Theta (\int_0^{z} ({k\cdot s + 1})^{-1}- \nC^{-1} \text{d}k)
 = \Theta \left(\frac{\log (zs)}{s}\right). 
\end{multline*}
Thus, it follows that $\Delta(d) = \Theta (\frac{\log z}{s})$.

\noindent\underline{(iii) $1 < \alpha< 2$:} 
In this case, as \eqref{eq:k}, 
$K = \Theta (1) $ and
\begin{multline*} 
\sum_{l=1}^{z}\sum^{\nC}_{i=1+(l-1)s}
\frac{1}{i^{\alpha}} =  \Theta \left(\int_0^{z-1}\int_{k\cdot
    s+1}^{\nC}\frac{1}{x^{\alpha}}\text{d}x \text{d}k \right)\cr
= \Theta (\int_0^{z-1} ({k\cdot s +1})^{1-\alpha}- \nC^{1-\alpha} \text{d}k)
 = \Theta \left(\frac{(zs)^{2-\alpha}}{s}\right).
\end{multline*}
Thus, it follows that $\Delta (d) = \Theta (\frac{(zs)^{2-\alpha}}{s}).$

\noindent\underline{(iv) $\alpha=1$:} 
In this case, $ K =\Theta \left( 1/ \log(\nC) \right)$ since
$1/K =\sum^{\nC}_{ i = 1} 1/i = \Theta \left(
\log (\nC) \right). $
The order of remaining terms is 
\begin{align*} 
\sum_{l=1}^{z}\sum^{\nC}_{i=1+(l-1)s}
\frac{1}{i} &=  \Theta \left(\int_0^{z-1}\int_{k\cdot
    s+1}^{\nC}\frac{1}{x}\text{d}x \text{d}k \right)
\cr &=  \Theta (\int_0^{z-1} \log (\nC)-\log (k\cdot s +1) \text{d}k)
\cr &= \Theta (z \log (\nC) - z \log (zs) + z ).
\end{align*}
Thus, it follows that
$\Delta (d) =  \Theta \left( \frac{z \log (\nC) - z \log (zs) + z }{\log(\nC)}\right).$



\noindent\underline{(v) $0< \alpha <1$:}
In this case,  $K=\Theta(\nC^{\alpha-1})$ and
we also find the order of remaining terms as follows:
\begin{align*} 
\sum_{l=1}^{z}\sum^{\nC}_{i=1+(l-1)s}
\frac{1}{i^{\alpha}} &= \Theta \left(\int_0^{z-1}\int_{k\cdot
    s+1}^{\nC}\frac{1}{x^{\alpha}}\text{d}x \text{d}k \right)
\cr &=\Theta (\int_0^{z-1}  \nC^{1-\alpha}- ({k\cdot s +1})^{1-\alpha} \text{d}k)
\cr &= \Theta \left( z \nC^{1-\alpha} -\frac{s^{1-\alpha} z^{2-\alpha}}{2-\alpha} \right).
\end{align*}
Since $sz \le \nC,$ we conclude that
$\Delta(d) = \Theta ( z ). $
Therefore, Theorem~\ref{thm:homo} for LBND follows.

\section{Heterogeneous Per-node
  Cache Size}\label{sec:non_equal_size}

\subsection{Motivation, Challenges and Model}

The study in Section~\ref{sec:equal_size} enables us to purely focus on
the impact of content popularity based caching on delay under the
assumption of equal per-node cache size. However, it may be possible to
gain more benefits by caching more contents at the caches that has more
geometric importance. Examples include the policy that assign more cache
budgets at the nodes with, e.g., high degrees or high access in
request routing. This section is devoted to quantifying such an impact
of heterogeneous cache sizing on delay. 


The heterogeneity in per-node cache size adds more challenges to
analysis. As done in Section \ref{sec:equal_size}, for homogenous cache
sizing, the probabilistic characteristics of random delay depends only
on the given routing distance, independent from their locations and the
routing path details. However, for the heterogeneous cache sizing, delay
depends on the routing path, not just the routing distance. Thus, it
seems inevitable to analyze the delay scaling laws in an
topology-dependent manner. 

\begin{figure}[h!]
  \centering
\includegraphics[width=0.6\columnwidth]{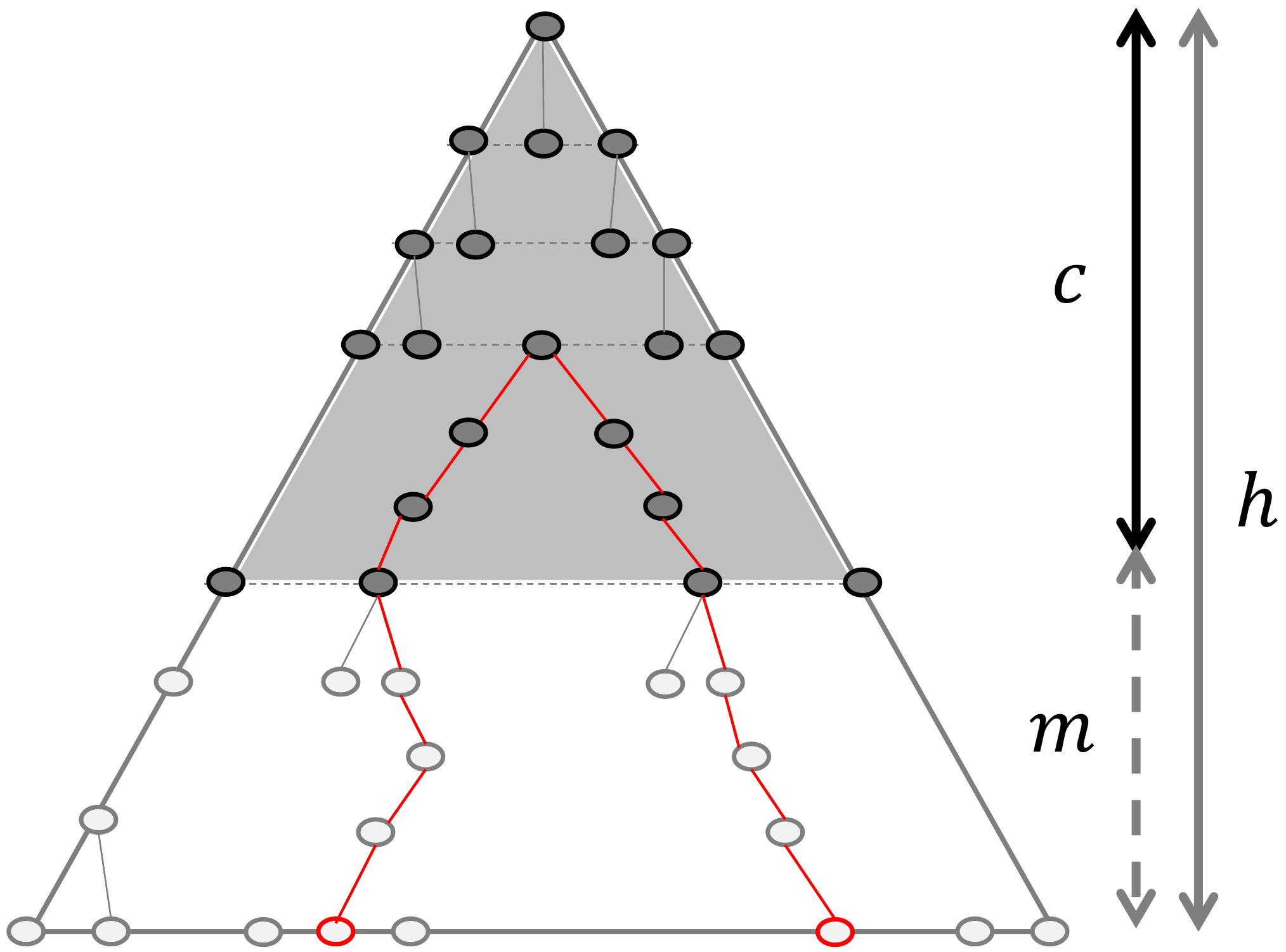}
  \caption{Regular spanning tree topology and BoW (Black or White) cache
    sizing policy, where the red line represents the shortest path between two nodes located in the bottom of the tree.} \label{fig:treetopology}
  \vspace{-0.2cm}
\end{figure}

\noindent{\bf \em Regular tree and BoW (Black or White) sizing policy.}
In this paper, we consider a cache network whose topology is a 
{\em $(r+1)$-regular spanning tree,} and {\em shortest-path} based routing, as
illustrated in Fig.~\ref{fig:treetopology}. The tree has total $n$ nodes
and $h$ layers, and each node has $r$ children except that the root has
$r+1$ children, such that every node has $r+1$ neighbors.  This enables us to cover a large class of popular
topologies, ranging from a line network to a star network, by simply
changing $r$ (e.g., a line for $r=1$).  We note that such tree
topologies have popularly  been used in P2P streaming systems
\cite{liu2010p2p, xu2009supporting} and several content routing
proposals in ICN \cite{bari2012survey}.  We comment that we assume
`perfect regularity' because of simplicity in analysis, and our work
can be readily extended to non-regular spanning trees.




There may be a large number of candidate cache sizing polices, out of
which we consider a very simple policy, called {\em BoW (Black or
  White)}, which partitions the entire nodes into nodes are cacheable
(black) and non-cacheable (white), respectively. In other words, the system-wide cache
budget is divided only among black nodes, and especially in BoW, 
the nodes only up-to the $c$-th layer become black, as seen in
Fig.~\ref{fig:treetopology}, where $c$ should be carefully chosen to
achieve low delay. Again, although not optimal, this simple
policy provides a lower bound on the gain from heterogeneous cache
sizing.








\subsection{Main Results}

\begin{table}[t!]
  \centering
  \tabcolsep=0.6mm
  \renewcommand{\arraystretch}{1.4} 
  \caption{Asymptotic delay with TPP-C and LBND policies when $B =
    \Theta (n),$ where $X=\min  [\log_r n , \nC ].$ $\Leftarrow$ means
    that the corresponding value is same as its left one. Similar
    meaning for $\Uparrow.$ \label{table:bw} }
  \vspace{-0.15cm}
  {\small
    \begin{tabular}{|c||c|c||c|c|} \hline
      &  \multicolumn{2}{c||}{{\em Homogeneous size  }}& \multicolumn{2}{c|}{{\em  Heterogeneous size  }}\\ \hline
      & TPP-C &LBND& TPP-C &LBND\\ \hline \hline
      $2<\alpha$& $\Theta(1)$ & $\Theta (1)$& $\Theta (1)$& $\Theta (1)$\\  \hline
      $\alpha=2$&  $ O(\log^2_r (X))$&  $ O(\log_r (X))$&  $ O(\log_r\log_r
      (X))$&  $\Leftarrow$ \\ \hline 
      $1< \alpha<2$& $O(X^{2-\alpha})$& $\Leftarrow$& $O( \log_r (X))$& $\Leftarrow$\\ \hline 
      
      $\alpha=1$&  \specialcell{$O ( \min [\log_r n,  $\\$\frac{\nC}{
          \log_r \nC}   ]  )$}&  $\Leftarrow$&  \specialcell{$O ( \log_r(\min [ n,  $\\$\nC ])  )$}&  $\Leftarrow$\\ \hline

      $0 < \alpha <1$& \specialcell{$O ( \min [
        \log_r n,  $\\$\nC   ]  )$}& $\Leftarrow$&  $\Uparrow$&  $\Leftarrow$\\ \hline
    \end{tabular}
  }
\end{table}

\begin{theorem}{Delay for heterogenous cache size}
\label{thm:sizing_popularity_content}
The average delay $\Delta$
scales as those in Table~\ref{table:bw}  for BoW cache sizing policy
and shortest-path request routing under the $(r+1)$-regular spanning tree topology.
\end{theorem}

The proof of Theorem~\ref{thm:sizing_popularity_content} is in Section~\ref{sec:proof-theor-refthm:hetero}. 
Here, we summarize the key proof techniques and the
interpretations of Theorem~\ref{thm:sizing_popularity_content}.
\smallskip
\begin{compactenum}[\em (a)]

\item In this topology, $h=\Theta (\log_r n)$ and for any fixed $c\ge
  1,$ the per-node cache size $b_v$ for each black node $v$ (i.e., nodes
  up-to the $c$-th layer) is $\Theta( \frac{B}{r^c} ).$ Let $m
  \triangleq h -c.$ The key lies in how to choose $c$ for small delay,
  as explained in what follows: First, note that  the delay is bounded by:
  $\Delta \le 2m + 2\Delta_{\text{black}}(m),$ where 
  $\Delta_{\text{black}}$ denotes the expected delay experienced in the ``black
  region'' whose bound can be computed by Theorem \ref{thm:homo}. 
  The best $m^\star$ that minimizes $2m + 2\Delta_{\text{black}}(m)$ is our
  interest, where $m^\star$ should be chosen such that the delays in the
  white and black regions are equal in the asymptotic sense, i.e.,   $m^\star = \Theta(\Delta_{\text{black}})(m^\star).$

\item TPP-C and LBND in heterogenous cache sizing achieve approximately $\log$-order delay reduction over
  those in homogeneous cache sizing. 
  In particular, TPP-C's delay becomes the same as LBND in all popularity bias. 
  
\item Caching gains due to cache sizing increase with the degree $r$ of
  the tree. 

\item For $\alpha \le 1$ (i.e., low popularity bias) and $\nC > \log_r
  n,$ where $\log_r n$ is the (worst-case) routing distance order in our tree topology, the delay becomes the
  same as the order of routing distance $d.$ This implies that caching
  cannot enhance delay.  

\item Recall that $m^\star = \Theta(\Delta_{\text{black}})(m^\star)$
  in $(a).$ From the results that the delay order decreases as $\alpha$
  increases, we can conclude that all cache nodes becomes useful,
  as the distribution of the content popularity is skewed more.


\end{compactenum}


\subsection{Proof of Theorem~\ref{thm:sizing_popularity_content}}\label{sec:proof-theor-refthm:hetero}

\noindent{\bf \em Proof for TPP-C.} 
First, note that $\Delta_{black}$ is the delay of TPP-C computable using
$Theorem~\ref{thm:homo}$ when the average distance is $c$ and per node cache size
$s=\Theta(\frac{B\cdot r^m}{n}).$ 
We again consider the following ranges of $\alpha$ separately.

\noindent\underline{(i) $2<\alpha$:} In this case, $\Delta_{black} =
\Theta (1)$ and
$\Delta = O ( \min\limits_{1 \le m \le h} 1+m ).$ In other words, the homogeneous
cache sizing is sufficient, i.e., $m=\Theta(1)$, and
$\Delta = \Theta(1).$

\smallskip
\noindent\underline{(ii) $\alpha = 2$:} In this case, $\Delta
(\bar{d}) = O\left(\frac{\log^2 (\min \{s\cdot \bar{d} , \nC \})}{s}\right)$ and
\begin{align*}
\Delta  &=  O \left( \min_{1 \le m \le h} m + \frac{n \cdot \log^2 ( \min \{ \frac{B\cdot r^m \cdot (h-m)}{n}, \nC \})}{B\cdot
  r^m}  \right)\\
&= O \left(
\log_r\left(\frac{n}{B}\log(\min\{\frac{B \log_r n}{n},\nC \})\right)\right),\end{align*}
where the minimum occurs when the order of $m$ is the same with the order of $\Delta .$

\smallskip
\noindent\underline{(iii) $1 < \alpha< 2$:} 
In this case, $\Delta (\bar{d}) = O\left(\frac{(\min
  \{s\cdot \bar{d}, \nC \})^{2-\alpha}}{s}\right)$ and
\begin{align*}
\Delta  &=  O \left( \min_{1 \le m \le h} m + \frac{n \cdot ( \min \{ \frac{B\cdot r^m \cdot (h-m)}{n}, \nC \})^{2-\alpha}}{B\cdot
  r^m}  \right)\cr
  &=  O \left(\log_r\left( \frac{n}{B}\min\{\log_r n,\nC \}\right)\right).
\end{align*}

\smallskip
\noindent\underline{(iv) $\alpha = 1$:} 
In this case, $\Delta
(\bar{d}) = O \left( \min \left\{ \bar{d},  \frac{\nC}{s \cdot \log \nC}   \right\}
\right)$ and
\begin{align*}
\Delta  &=  O \left( \min_{1 \le m \le h} m + \min \left\{ h-m, \frac{n \cdot \nC   }{B\cdot
  r^m \cdot \log \nC} \right\} \right)\cr
  &=  O \left(\log_r\left(\min\{ n , \frac{n}{B} \cdot\nC \}\right)\right).
\end{align*}

\smallskip
\noindent\underline{(v) $1> \alpha >0 $:} In this case, $\Delta (\bar{d}) = O \left( \min \left\{ \bar{d},  \frac{\nC}{s}   \right\} \right)$ and
\begin{align*}
\Delta  &=  O \left( \min_{1 \le m \le h} m + \min \left\{ h-m, \frac{n \cdot \nC   }{B\cdot
  r^m } \right\} \right)\\
  &=  O \left(\log_r\left(\min\{ n , \frac{n}{B} \cdot\nC \}\right)\right).
\end{align*}
Therefore, Theorem~\ref{thm:sizing_popularity_content} for TPP-C follows.


\smallskip
\noindent{\bf \em Proof for LBND.}
First, observe that that in Theorem~\ref{thm:homo}, the delay order of LBND is the same
with that of TPP-C except the case $\alpha=2.$ Thus, we
only consider $\alpha=2$ and in this case,
$\Delta (\bar{d}) = \Theta\left(\frac{\log (\min
  \{s\cdot \bar{d} , \nC \})}{s}\right)$ due to Theorem~\ref{thm:homo}.
When we put $\bar{d} = c$
and $s = \Theta(\frac{B\cdot r^m}{n}),$ we have
\begin{align*}
\Delta  &=  O \left( \min_{1 \le m \le h} m + \frac{n \cdot \log ( \min \{ \frac{B\cdot r^m \cdot (h-m)}{n}, \nC \})}{B\cdot
  r^m}  \right)\\
&= O \left(
\log_r\left(\frac{n}{B}\log(\min\{\frac{B \log_r n}{n},\nC \})\right)\right).
\end{align*}
Therefore, Theorem~\ref{thm:sizing_popularity_content} for LBND follows.


\section{Simulation Results}
\label{sec:simulation}

\begin{figure}[t!]  
  \centering
  \includegraphics[width=0.99\columnwidth]{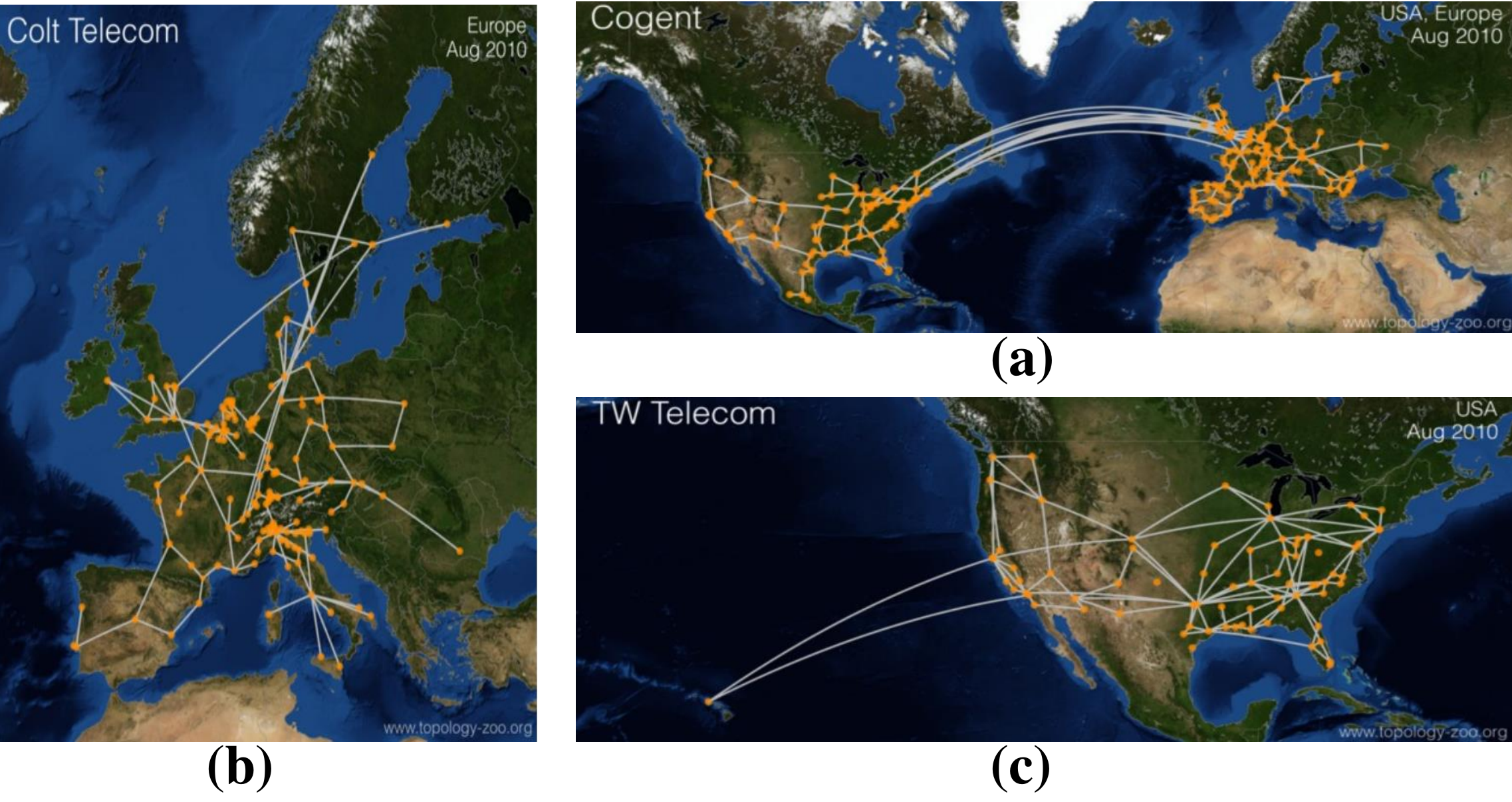}
  \caption{AS topologies of (a) Cogent (Europe-USA), (b) Colt Telecom (Europe), and (c) TW Telecom (USA) \cite{topologyzoo}} \label{fig:network_picture}
  \vspace{0.2cm}
\end{figure}

In this section, we verify the results in Theorems \ref{thm:homo} with scenarios I (line topology) and II (three AS topologies; Cogent (USA-Europe), Colt Telecom (Europe),
and TW Telecom (USA) from \cite{topologyzoo}, as seen in
Fig.~\ref{fig:network_picture}), and identify the effects of BoW cache sizing policy in Theorem \ref{thm:sizing_popularity_content} with scenario III (tree topology in Fig.~\ref{fig:treetopology}). In scenario III, we consider that all requests arrive at one of nodes located in bottom of the tree, and all content servers are attached to the top node. Thus, all shortest path, from a requested node to the server, pass through the cacheable (black) and non-cacheable (white) regions in BoW cache sizing policy , as we mentioned in Section \ref{sec:non_equal_size}. We conducted an off-line processing to understand topologies,
presented in Table~\ref{table:simulation}, whose features are
sufficiently heterogeneous without the average degree.  

    \begin{table}[h!]
    \centering
  \renewcommand{\tabcolsep}{0.7mm} 
  \renewcommand{\arraystretch}{1.3} 
   \caption{Simulation environments in scenarios I, II, and III \label{table:simulation} }
   {\small
 \begin{tabular}{|c||c|ccc|c|} \hline 
   Scenario  & I &\multicolumn{3}{c|}{ II} & III  \\ \hline 
  Topology &Line     &Cogent& Colt Tel. &Tw Tel.& Tree\\ \hline 
  $N$    	& 200	&197& 153& 76 & 98302\\ \hline
  Average degree & 2 	&2.49& 2.50& 3.08 &3 \\ \hline
  $\bar{d}$&66.33		&10.40& 8.24& 3.21 		&15\\ \hline
  $\nC$, $s$	& 400, 50	&\multicolumn{3}{c|}{3000, 5}	&3000, 5\\ \hline
  $\hat{i}=\min[s\cdot \bar{d},\nC]$	&400	&50 & 40 & 15	&75\\ \hline
     \end{tabular}
     }
 \end{table} 

\begin{figure}[t!] 
\centering
\includegraphics[width=0.7\columnwidth]{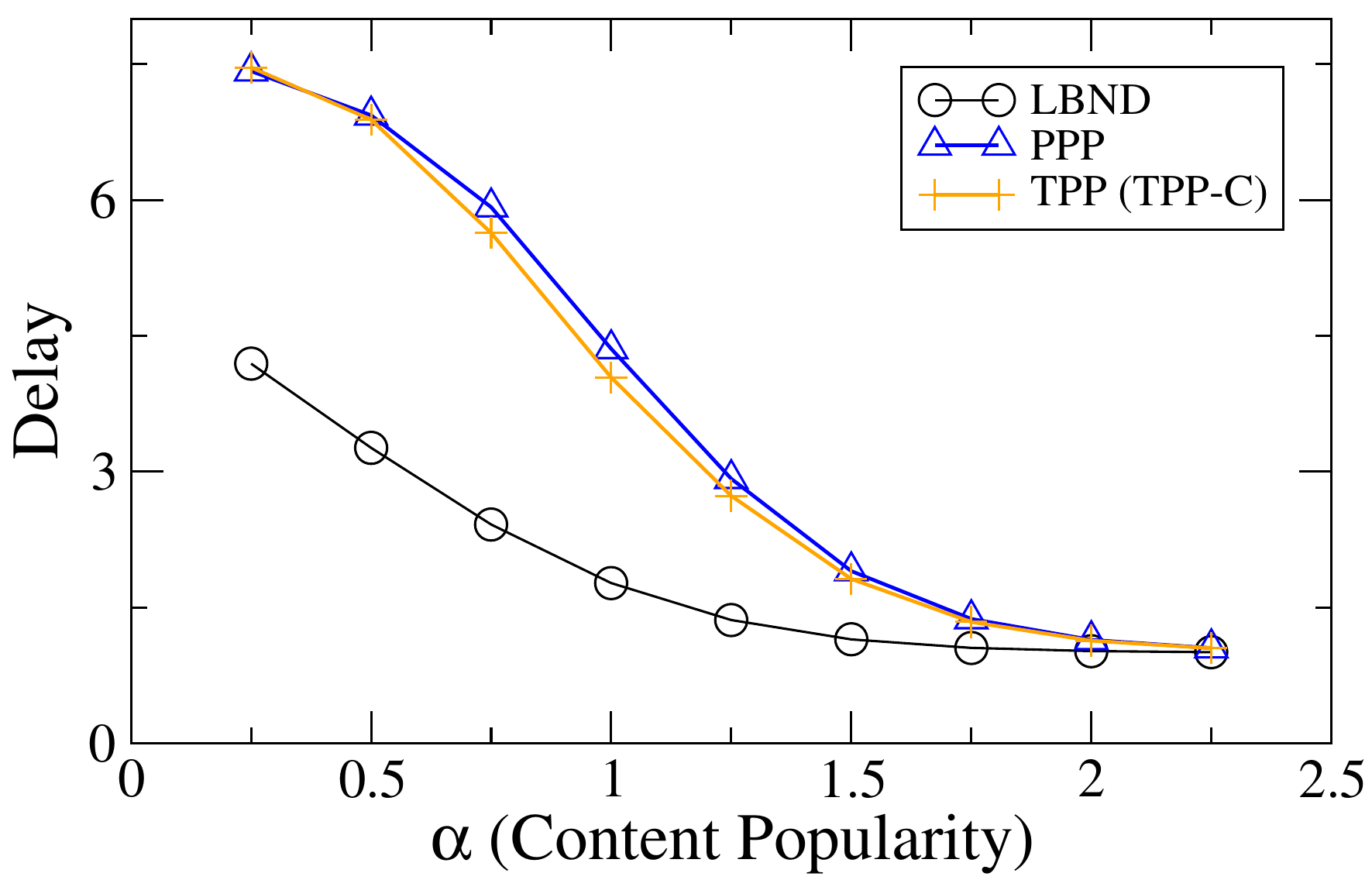} \\
\vspace{0.2cm}
\caption{Delay performance for line topology with $s=50$ in scenario I: the average delays of TPP (TPP-C) is less than or equal to that of PPP.} \label{fig:line}
\end{figure}

\begin{figure*}[t!] 
\centering
{\small
\begin{tabular}{ccc}
\includegraphics[width=0.65\columnwidth]{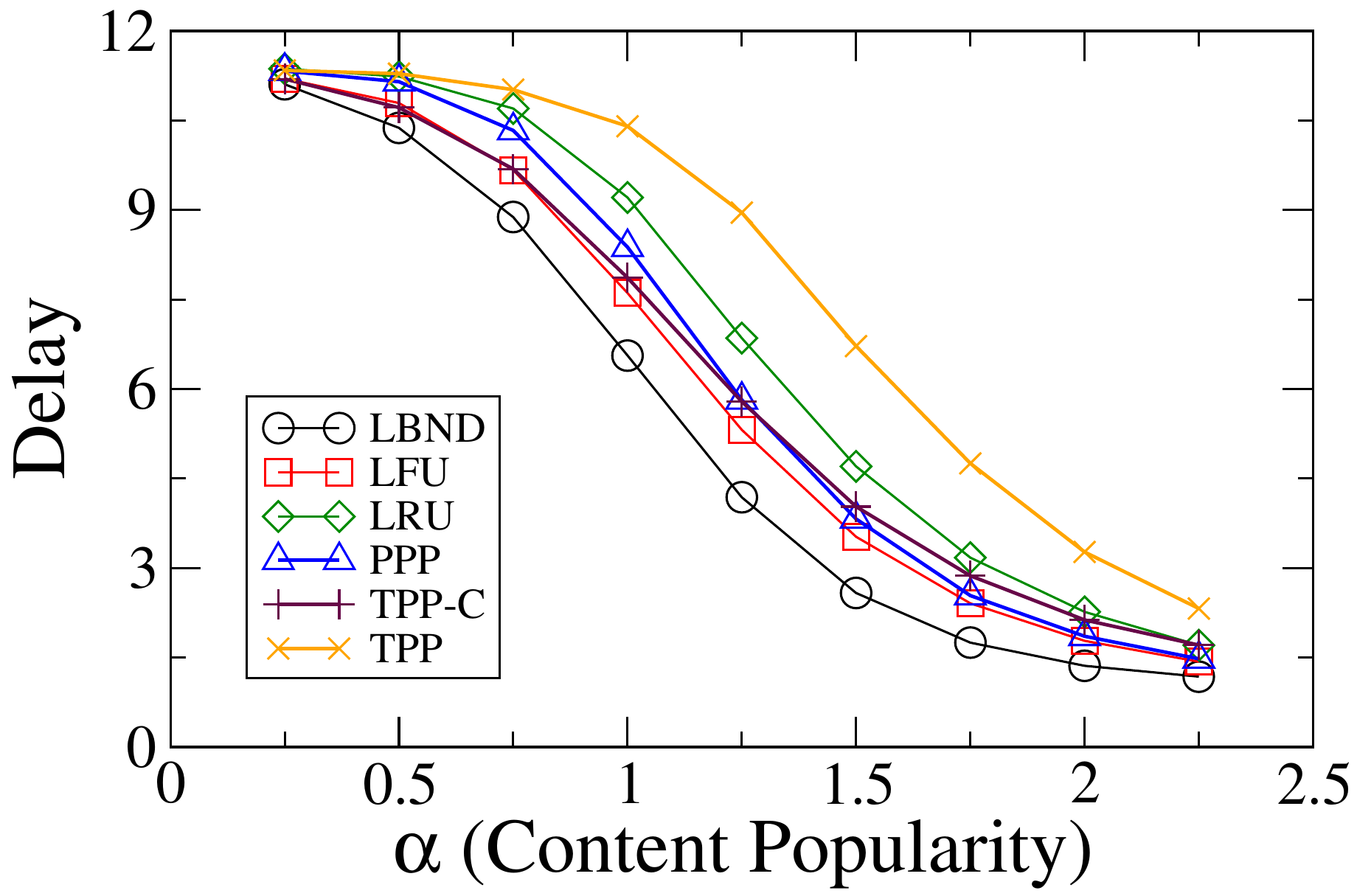}&
\includegraphics[width=0.65\columnwidth]{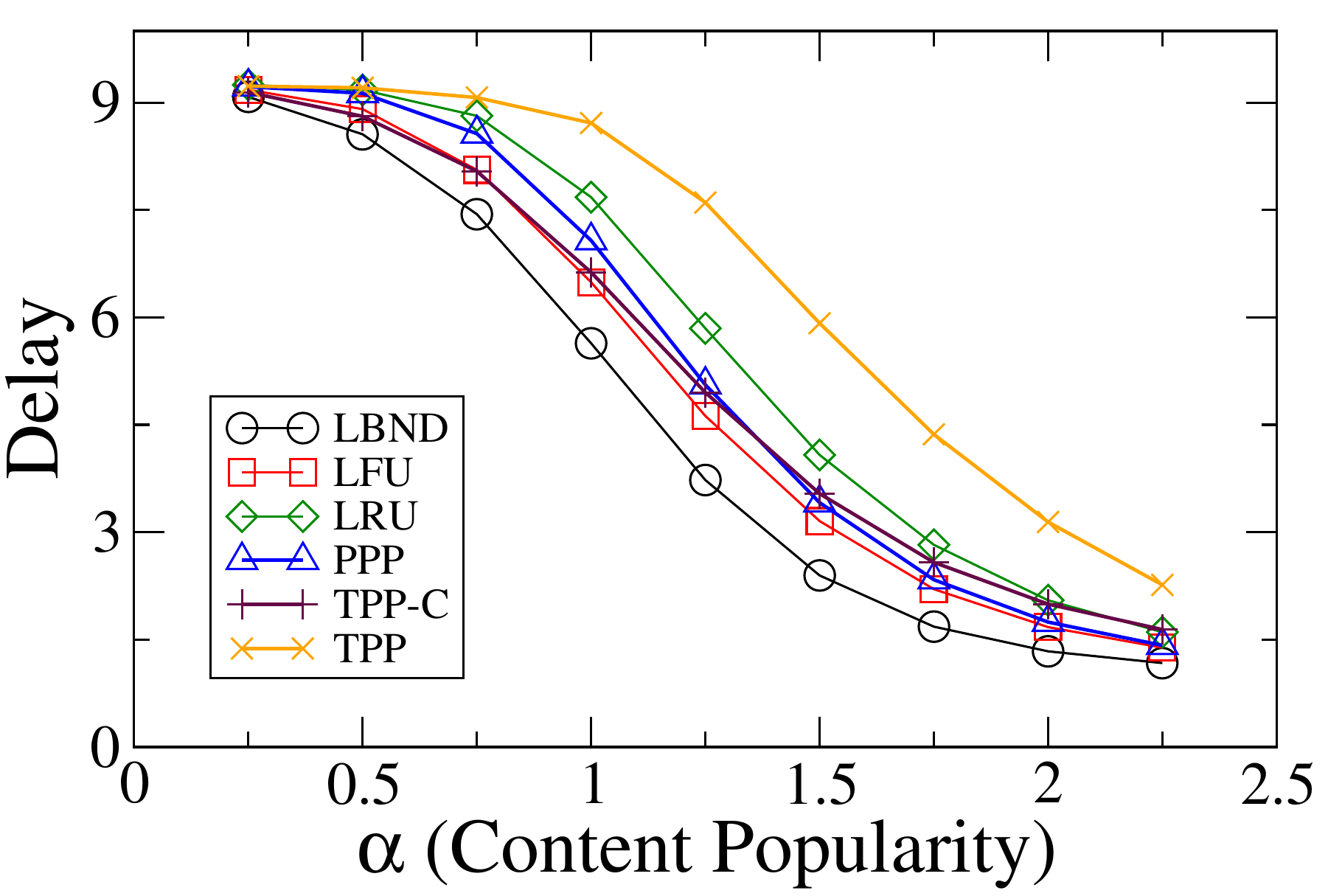}&
\includegraphics[width=0.65\columnwidth]{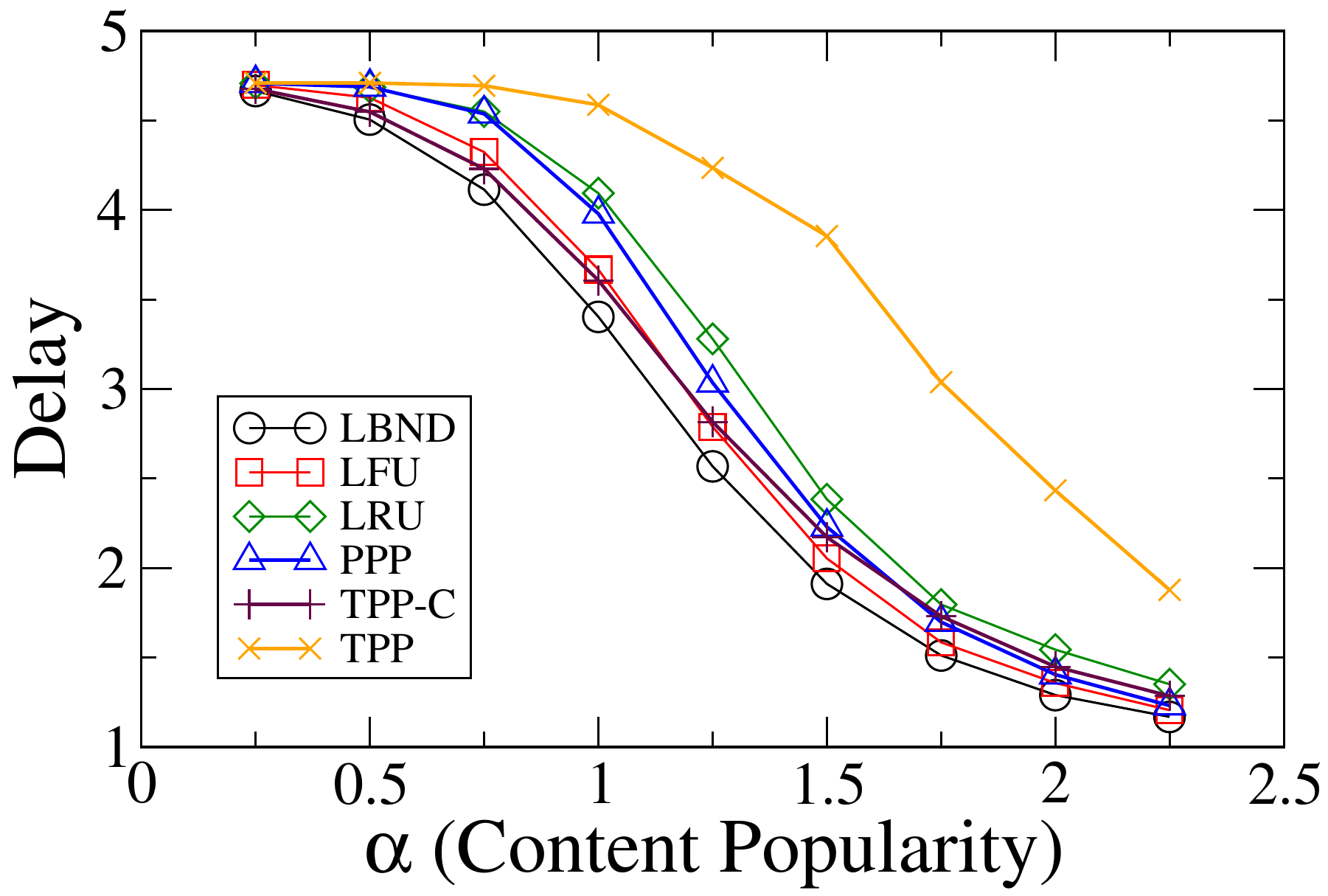} \\
(a) Cogent & (b) Colt & (c) TW
\end{tabular}}
\vspace{0.2cm}
\caption{Delay performance for three AS topologies in scenario II: PPP and TPP-C's
  delay is slightly larger and smaller than LFU and LRU,
  respectively. Thus, our analysis in this paper is useful to predicting
  the performance of cache networks. } \label{fig:as}
\vspace{-0.4cm}
\end{figure*}

\begin{figure}[t!]  
\centering
\includegraphics[width=0.7\columnwidth]{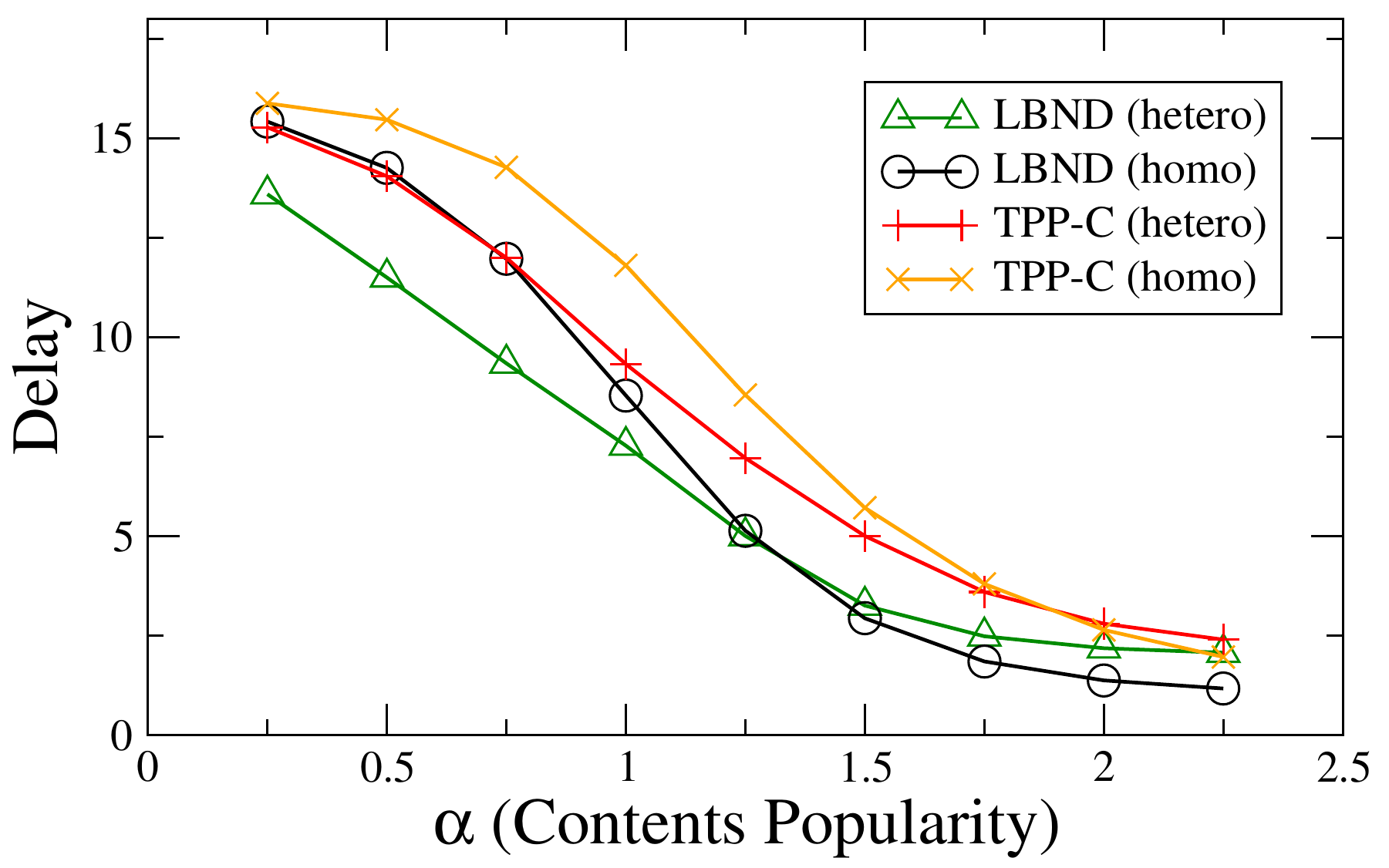} \\
\vspace{0.2cm}
\caption{Delay performance for tree topology in scenario III: the average delay with heterogeneous caching is lower than that with homogeneous caching when $\alpha<2.$} \label{fig:tree}
\end{figure}

We construct the simulation environment such that $s \cdot \bar{d} \ll \nC$ except scenario I, based on the recent trend of explosive increase in the number of contents. To get simulation
results, we perform 10 times of random instances during 100000 slots. 
For each test, we first place content servers uniformly at random, and 
a content request arrives at a cache with probability
0.5 at the beginning of the time slot, and unresolved requests are
forwarded to the next cache under the shortest path routing. 
Figures~\ref{fig:line}, ~\ref{fig:as} and ~\ref{fig:tree} show the results of various
caching policies over scenarios I, II and III, respectively. We test not only static strategies
introduced in this work, but also dynamic caching strategies LFU and
LRU. 



\noindent{\bf \em (i) Comparison of static strategies, PPP, TPP, and TPP-C.}
As indicated in Theorem \ref{thm:homo}, the average delays of TPP (TPP-C) is slightly less than or equal to that of PPP when $s \cdot \bar{d} \gg \nC$ as illustrated in Fig.~\ref{fig:line}. On the other hand, we observe that the average delays of PPP and TPP-C are far better than that of TPP when $s \cdot \bar{d} \ll \nC$ as shown in Fig.~\ref{fig:as}. Moreover, TPP-C policy
generally outperforms other policies for any topology and $\alpha.$
Especially, in comparison between TPP-C and PPP, as stated in Theorem
\ref{thm:homo} (and Table~\ref{table:total}), their delay difference
becomes higher for low content popularity bias. 


\smallskip
\noindent {\bf \em (ii) Comparison with dynamic strategies LFU and LRU.} 
In this work, we provide asymptotic analysis with static policies
instead of real, dynamic cache replacement policies such as LFU and LRU,
operating without any prior knowledge about the content popularity
distribution.  However, the average delays with PPP and TPP-C are less
than or equal to that with LRU policy and slightly greater than that
with LFU policy, as shown in Fig.~\ref{fig:as}, basically meaning that
our results can be useful to predict the delay performance of real cache
networks.

\smallskip
\noindent {\bf \em (iii) Comparison with homogeneous caching and BoW cache sizing policy for LBND and TPP-C.} 
Fig.~\ref{fig:tree} shows the effectiveness of BoW cache sizing policy as we verified in Theorem \ref{thm:sizing_popularity_content}. We obtain the average delay with homogeneous caching when the nodes up to the 15-th layer are cacheable (Black). On the other hand, the average delay with heterogeneous caching is the minimum value among the average delays of four cases when the nodes up to the 11-th, 12-th, 13-th, and 14-th layers are cacheable, respectively. For $\alpha \ge 2$, the average delay of TPP-C with heterogeneous caching is slightly greater than that with homogeneous caching. Since the request popularity is centralized in few contents, there is no gain from the heterogeneous caching. On the other hand, for $\alpha <2$, the average delay of TPP-C is reduced by heterogeneous caching policy.


\vspace{0.5cm}
\vspace{0.2cm}

\section{Conclusion}
  In this paper, we performed asymptotic analysis of the delay
  performance of large-scale cache networks. We focused on quantitively
  understanding the relation between content popularity and delay
  performance as well as the impact of heterogeneity in terms of “node
  importance”. We first studied the asymptotic delay performance of
  cache networks under homogenous per-node cache budget. We showed that
  there are small and large gains in popularity-based content placements
  when contents are highly homogenous and heterogeneous,
  respectively. Second, we showed the asymptotic delay performance of
  cache networks under heterogeneous per-node cache budget and the
  caching gain incurred by heterogeneous cache sizing using nodes’
  geometric importances increases.


\balance
\renewcommand{\baselinestretch}{0.95}
\bibliographystyle{IEEEtran}


\begin{thebibliography}{10}
\providecommand{\url}[1]{#1}
\csname url@samestyle\endcsname
\providecommand{\newblock}{\relax}
\providecommand{\bibinfo}[2]{#2}
\providecommand{\BIBentrySTDinterwordspacing}{\spaceskip=0pt\relax}
\providecommand{\BIBentryALTinterwordstretchfactor}{4}
\providecommand{\BIBentryALTinterwordspacing}{\spaceskip=\fontdimen2\font plus
\BIBentryALTinterwordstretchfactor\fontdimen3\font minus
  \fontdimen4\font\relax}
\providecommand{\BIBforeignlanguage}[2]{{%
\expandafter\ifx\csname l@#1\endcsname\relax
\typeout{** WARNING: IEEEtran.bst: No hyphenation pattern has been}%
\typeout{** loaded for the language `#1'. Using the pattern for}%
\typeout{** the default language instead.}%
\else
\language=\csname l@#1\endcsname
\fi
#2}}
\providecommand{\BIBdecl}{\relax}
\BIBdecl

\bibitem{DONA}
T.~Koponen, M.~Chawla, B.-G. Chun, A.~Ermolinskiy, K.~H. Kim, S.~Shenker, and
  I.~Stoica, ``A data-oriented (and beyond) network architecture,'' in
  \emph{Proc. ACM SIGCOMM}, 2007.

\bibitem{NDN}
L.~Zhang, D.~Estrin, J.~Burke, V.~Jacobson, J.~D. Thornton, D.~K. Smetters,
  B.~Zhang, G.~Tsudik, D.~Massey, C.~Papadopoulos \emph{et~al.}, ``Named data
  networking {(NDN)} project,'' \emph{Relat{\'o}rio T{\'e}cnico NDN-0001, Xerox
  Palo Alto Research Center-PARC}, 2010.

\bibitem{CCN}
V.~Jacobson, D.~K. Smetters, J.~D. Thornton, M.~F. Plass, N.~H. Briggs, and
  R.~L. Braynard, ``{Networking named content},'' in \emph{Proc. ACM CoNext},
  2009.

\bibitem{4WARD}
N.~Niebert, S.~Baucke, I.~El-Khayat, M.~Johnsson, B.~Ohlman, H.~Abramowicz,
  K.~Wuenstel, H.~Woesner, J.~Quittek, and L.~Correia, ``{The way {4WARD} to
  the creation of a future {I}nternet},'' in \emph{Proc. IEEE PIMRC}, 2008.

\bibitem{PSIRP/PURSUIT}
N.~Fotiou, G.~C. Polyzos, P.~Nikander, and D.~Trossen, ``{Developing
  information networking further: from {PSIRP} to {PURSUIT}},'' in \emph{Proc.
  ICST Broadnets}, 2010.

\bibitem{COMET}
G.~Garcia, A.~Beben, F.~J. Ramon, A.~Maeso, I.~Psaras, G.~Pavlou, N.~Wang,
  J.~Sliwinski, S.~Spirou, S.~Soursos, and E.~Hadjioannou, ``{{COMET}: content
  mediator architecture for content-aware networks},'' in \emph{In Proc. Future
  Network and Mobile Summit}, 2011.

\bibitem{che2001analysis}
H.~Che, Z.~Wang, and Y.~Tung, ``Analysis and design of hierarchical {W}eb
  caching systems,'' in \emph{Proc. Infocom}, 2001.

\bibitem{che2002hierarchical}
H.~Che, Y.~Tung, and Z.~Wang, ``Hierarchical {W}eb caching systems: modeling,
  design and experimental results,'' \emph{IEEE Journal on Selected Areas in
  Communications}, vol.~20, no.~7, pp. 1305--1314, 2002.

\bibitem{dan1990approximate}
A.~Dan and D.~Towsley, ``An approximate analysis of the {LRU} and {FIFO} buffer
  replacement schemes,'' \emph{Performance Evaluation Review}, vol.~18, no.~1,
  pp. 143--152, 1990.

\bibitem{jelenkovic1999asymptotic}
P.~Jelenkovi{\'c}, ``Asymptotic approximation of the move-to-front search cost
  distribution and least-recently-used caching fault probabilities,'' \emph{The
  Annals of Applied Probability}, vol.~9, no.~2, pp. 430--464, 1999.

\bibitem{psaras2011modelling}
I.~Psaras, R.~G. Clegg, R.~Landa, W.~K. Chai, and G.~Pavlou, ``Modelling and
  evaluation of {CCN}-caching trees,'' in \emph{Proc. NETWORKING}.\hskip 1em
  plus 0.5em minus 0.4em\relax Springer, 2011, pp. 78--91.

\bibitem{gitzenis2012asymptotic}
S.~Gitzenis, G.~S. Paschos, and L.~Tassiulas, ``Asymptotic laws for content
  replication and delivery in wireless networks,'' in \emph{Proc. Infocom},
  2012.

\bibitem{jelenkovic2005near}
P.~R. Jelenkovic, X.~Kang, and A.~Radovanovic, ``Near optimality of the
  discrete persistent access caching algorithm,'' in \emph{Proc. International
  Conference on Analysis of Algorithms}, 2005.

\bibitem{rodriguez2001analysis}
P.~Rodriguez, C.~Spanner, and E.~W. Biersack, ``Analysis of {W}eb caching
  architectures: hierarchical and distributed caching,'' \emph{IEEE
  Transactions on Networking}, vol.~9, no.~4, pp. 404--418, 2001.

\bibitem{fricker2012versatile}
C.~Fricker, P.~Robert, and J.~Roberts, ``A versatile and accurate approximation
  for {LRU} cache performance,'' in \emph{Proc. ITC}, 2012.

\bibitem{borst2010distributed}
S.~Borst, V.~Gupta, and A.~Walid, ``Distributed caching algorithms for content
  distribution networks,'' in \emph{Proc. IEEE Infocom}, 2010.

\bibitem{rosensweig2010approximate}
E.~Rosensweig, J.~Kurose, and D.~Towsley, ``Approximate models for general
  cache networks,'' in \emph{Proc. Infocom}, 2010.

\bibitem{gallo2012performance}
M.~Gallo, B.~Kauffmann, L.~Muscariello, A.~Simonian, and C.~Tanguy,
  ``Performance evaluation of the random replacement policy for networks of
  caches,'' in \emph{Proc. ACM SIGMETRICS}, 2012.

\bibitem{CSsize}
D.~Rossi and G.~Rossini, ``{On sizing {CCN} content stores by exploiting
  topological information},'' in \emph{Proc. IEEE NOMEN}, 2012.

\bibitem{muscariello2011bandwidth}
L.~Muscariello, G.~Carofiglio, and M.~Gallo, ``Bandwidth and storage sharing
  performance in information centric networking,'' in \emph{Proc. ACM SIGCOMM
  workshop on Information-centric networking}, 2011.

\bibitem{carofiglio2011modeling}
G.~Carofiglio, M.~Gallo, L.~Muscariello, and D.~Perino, ``Modeling data
  transfer in content-centric networking,'' in \emph{Proc. ITC}, 2011.

\bibitem{azimdoost2012throughput}
B.~Azimdoost, C.~Westphal, and H.~R. Sadjadpour, ``On the throughput capacity
  of information-centric networks,'' \emph{arXiv preprint arXiv:1210.1185},
  2012.

\bibitem{chung2002average}
F.~Chung and L.~Lu, ``The average distances in random graphs with given
  expected degrees,'' vol.~99, no.~25.\hskip 1em plus 0.5em minus 0.4em\relax
  National Acad Sciences, 2002, pp. 15\,879--15\,882.

\bibitem{draief2010epidemics}
M.~Draief and L.~Massouli, \emph{Epidemics and rumours in complex
  networks}.\hskip 1em plus 0.5em minus 0.4em\relax Cambridge University Press,
  2010.

\bibitem{liu2010p2p}
S.~Liu, M.~Chen, S.~Sengupta, M.~Chiang, J.~Li, and P.~A. Chou, ``{P2P}
  streaming capacity under node degree bound,'' in \emph{Proc. IEEE Distributed
  Computing Systems}, 2010.

\bibitem{xu2009supporting}
T.~Xu, J.~Chen, W.~Li, S.~Lu, Y.~Guo, and M.~Hamdi, ``Supporting {VCR}-like
  operations in derivative tree-based {P2P} streaming systems,'' in \emph{Proc.
  ICC}, 2009.

\bibitem{bari2012survey}
M.~Bari, S.~Rahman~Chowdhury, R.~Ahmed, R.~Boutaba, and B.~Mathieu, ``A survey
  of naming and routing in information-centric networks,'' \emph{IEEE
  Communications Magazine}, vol.~50, no.~12, pp. 44--53, 2012.

\bibitem{topologyzoo}
``{The {I}nternet topology zoo},'' http://www.topology-zoo.org/dataset.html.

\end{thebibliography}






\end{document}